\documentclass[twocolumn,showpacs,aps,prx,groupedaddress,superscriptaddress]{revtex4-2}

\usepackage{mathrsfs}
\usepackage{amssymb, amsbsy, amsmath, latexsym, dsfont, array, layout, mathrsfs, color}

\usepackage{mathtools}

\usepackage{amsmath,amssymb}
\usepackage{bm}
\usepackage{graphicx}
\usepackage{braket,dsfont}
\usepackage{color}
\usepackage[10pt]{moresize}
\usepackage{mathrsfs}
\usepackage{multirow}
\usepackage{hyperref}
\usepackage{changes}
\usepackage{mathdots}
\let\oldcaption\caption
\renewcommand{\caption}{\sffamily \oldcaption}

\newcommand{\ketbra}[2]{\ket{#1}\!\!\bra{#2}}

\usepackage{amsmath,amsfonts,amssymb}
\usepackage{wrapfig}
\usepackage{graphicx}
\usepackage{bbm}
\usepackage[normalem]{ulem}
\usepackage{color}

\begin{document}

\title{Quantification of Photon Fusion \\ for Genuine Multiphoton Quantum Correlations}\date{\today}

\author{Sheng-Yan Sun}
\affiliation{Department of Engineering Science, National Cheng Kung University, Tainan 70101, Taiwan}
\affiliation{Center for Quantum Frontiers of Research $\&$ Technology, National Cheng Kung University, Tainan 70101, Taiwan}

\author{Yu-Cheng Li}
\affiliation{Department of Engineering Science, National Cheng Kung University, Tainan 70101, Taiwan}
\affiliation{Center for Quantum Frontiers of Research $\&$ Technology, National Cheng Kung University, Tainan 70101, Taiwan}

\author{Shih-Hsuan Chen}
\affiliation{Department of Engineering Science, National Cheng Kung University, Tainan 70101, Taiwan}
\affiliation{Center for Quantum Frontiers of Research $\&$ Technology, National Cheng Kung University, Tainan 70101, Taiwan}

\author{Kuan-Jou Wang}
\affiliation{Department of Engineering Science, National Cheng Kung University, Tainan 70101, Taiwan}
\affiliation{Center for Quantum Frontiers of Research $\&$ Technology, National Cheng Kung University, Tainan 70101, Taiwan}

\author{Ching-Jui Huang}
\affiliation{Department of Engineering Science, National Cheng Kung University, Tainan 70101, Taiwan}
\affiliation{Center for Quantum Frontiers of Research $\&$ Technology, National Cheng Kung University, Tainan 70101, Taiwan}

\author{Tung-Ju Tsai}
\affiliation{Department of Engineering Science, National Cheng Kung University, Tainan 70101, Taiwan}
\affiliation{Center for Quantum Frontiers of Research $\&$ Technology, National Cheng Kung University, Tainan 70101, Taiwan}

\author{Wei-Ting Kao}
\affiliation{Department of Engineering Science, National Cheng Kung University, Tainan 70101, Taiwan}
\affiliation{Center for Quantum Frontiers of Research $\&$ Technology, National Cheng Kung University, Tainan 70101, Taiwan}

\author{Tzu-Liang Hsu}
\affiliation{Department of Engineering Science, National Cheng Kung University, Tainan 70101, Taiwan}
\affiliation{Center for Quantum Frontiers of Research $\&$ Technology, National Cheng Kung University, Tainan 70101, Taiwan}
\affiliation{Physics Department, Blackett Laboratory, Imperial College London, Prince Consort Road, SW7 2BW, United Kingdom}

\author{Che-Ming Li}
\email{cmli@mail.ncku.edu.tw}
\affiliation{Department of Engineering Science, National Cheng Kung University, Tainan 70101, Taiwan}
\affiliation{Center for Quantum Frontiers of Research $\&$ Technology, National Cheng Kung University, Tainan 70101, Taiwan}
\affiliation{Center for Quantum Science and Technology, Hsinchu 30013, Taiwan}

\begin{abstract}
Fusing photon pairs creates an arena where indistinguishability can exist between two two-photon amplitudes contributing to the same joint photodetection event. This two-photon interference has been extensively utilized in creating multiphoton entanglement, from passive to scalable generation, from bulk-optical to chip-scale implementations. While significant, no experimental evidence exists that the full capability of photon fusion can be utterly quantified like a quantum entity. Herein, we demonstrate the first complete capability quantification of experimental photon fusion. Our characterization faithfully measures the whole abilities of photon fusion in the experiment to create and preserve entangled photon pairs. With the created four- and six-photon entangled states using spontaneous parametric down-conversion entanglement sources, we show that capability quantification provides a faithful assessment of interferometry for generating genuine multiphoton entanglement and Einstein-Podolsky-Rosen steering. These results reveal a practical diagnostic method to benchmark photon fusion underlying the primitive operations in general quantum photonics devices and networks.
\end{abstract}
\maketitle


\begin{figure}[t]
\includegraphics[width=8cm]{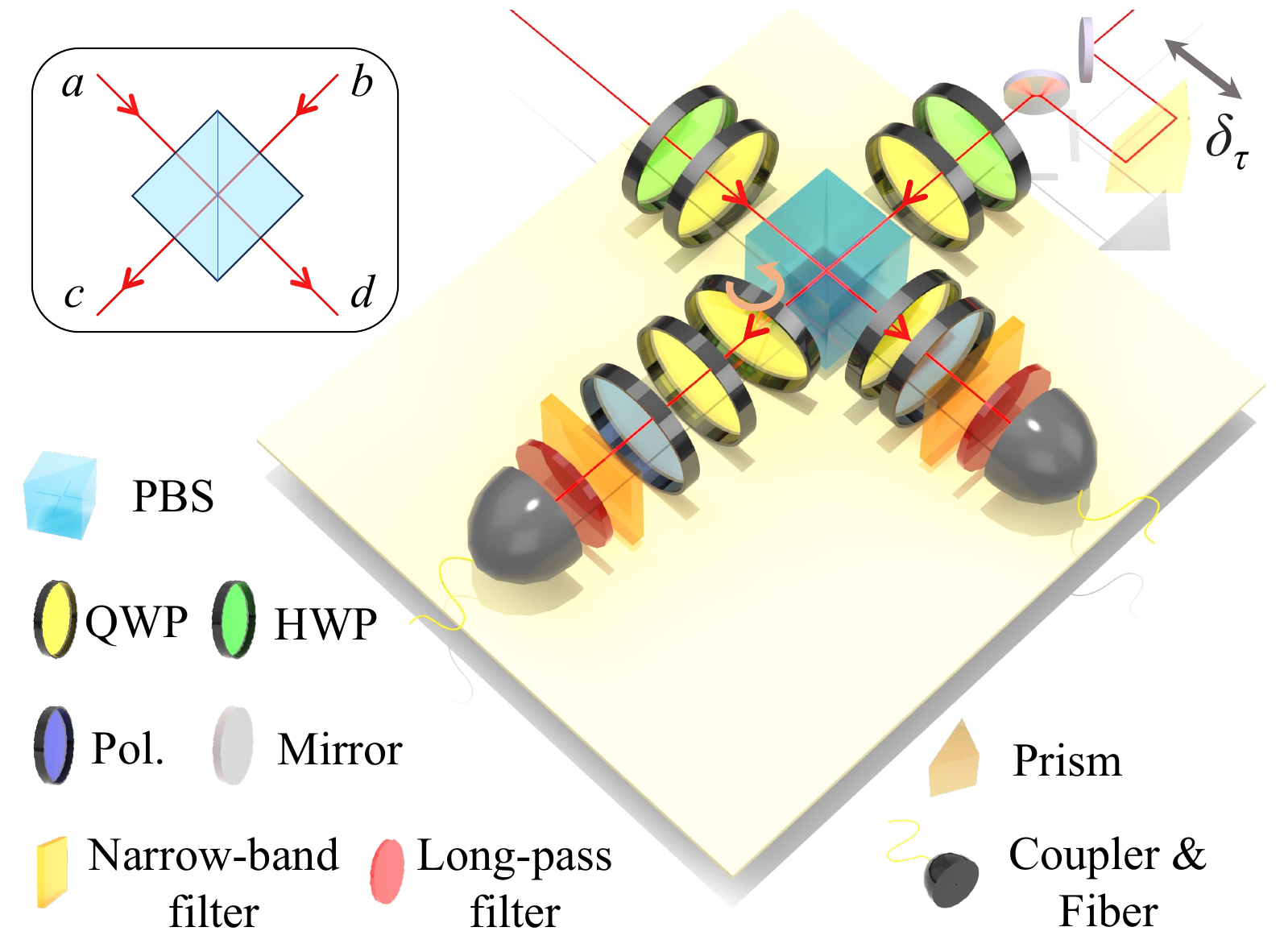}
\caption{Experimental setup of photon fusion unit. Two individual photons in the spatial modes $a$ and $b$ are input into a polarizing beam splitter (PBS) and post-selected in the spatial modes $c$ and $d$. To temporally and spatially overlap at the PBS, the path of photon $b$ is adjusted by fine-tuning the delay $\delta_{\tau}$ with a nanopositioning stage. Half wave plate (HWP) and quarter wave plate (QWP) are used to prepare different inputs for process tomography. A QWP and a polarizer (Pol.) is combined for projective measurements. The relative phase between the two-photon amplitudes is adjuted by properly rotating an additional QWP around its vertically-oriented fast axis in mode $c$. Photons were selected by the narrow-band filter and long-pass filter, collected by fiber couplers, and detected by silicon avalanche single-photon detectors; coincidences of single photons in modes $c$ and $d$ are recorded by an FPGA (field programmable gate array) based coincidence unit (not shown).}
\label{Fig1}
\end{figure}

\section{Introduction}

Photon fusion~\cite{Pan98,Pan01,Pan03} is a nonclassical process of superposing (or fusing) two-photon amplitudes of different yet indistinguishable alternatives of a joint photodetection event. This leads to entanglement in the polarization degree of freedom between a photon pair~\cite{Browne05,Bodiya06,Lu07}. Since its experimental implementation only demands moderate linear passive optical elements, beam-path adjustment techniques, single-photon detection, and coincidence counting (Fig.~\ref{Fig1}), this two-photon interferometry has served as a building block for multiphoton entanglement in bulk optics~\cite{Pan12,Wang16,Zhong18} and on-chip waveguide quantum circuits~\cite{Adcock19,Llewellyn20,Wang20}, including the scalable creation by active feed-forward and multiplexing~\cite{MeyerScott22}. 

These advances make multiphoton entanglement widely used to investigate quantum mechanics~\cite{Pan00,Lu20,Wu22} and information~\cite{Walther05,Lu09,Yao12}. Moreover, genuine multipartite entanglement powers quantum networks' correlation, coordination, and security~\cite{Kimble08,Ritter12,Wehner18}. When genuine multiphoton entanglement is created to possess stronger quantum correlations, such as genuine multipartite Einstein-Podolsky-Rosen (EPR) steerability~\cite{He13,Li15,Armstrong15,Cavalcanti15}, multiphoton quantum correlations provide flexibility of entanglement shared by untrusted network nodes~\cite{Lu20,Kao23}. With such a central role, the photon fusion process's experimental performance ultimately decides these applications' reliability.

\begin{figure*}[t]
\includegraphics[width=14.3cm]{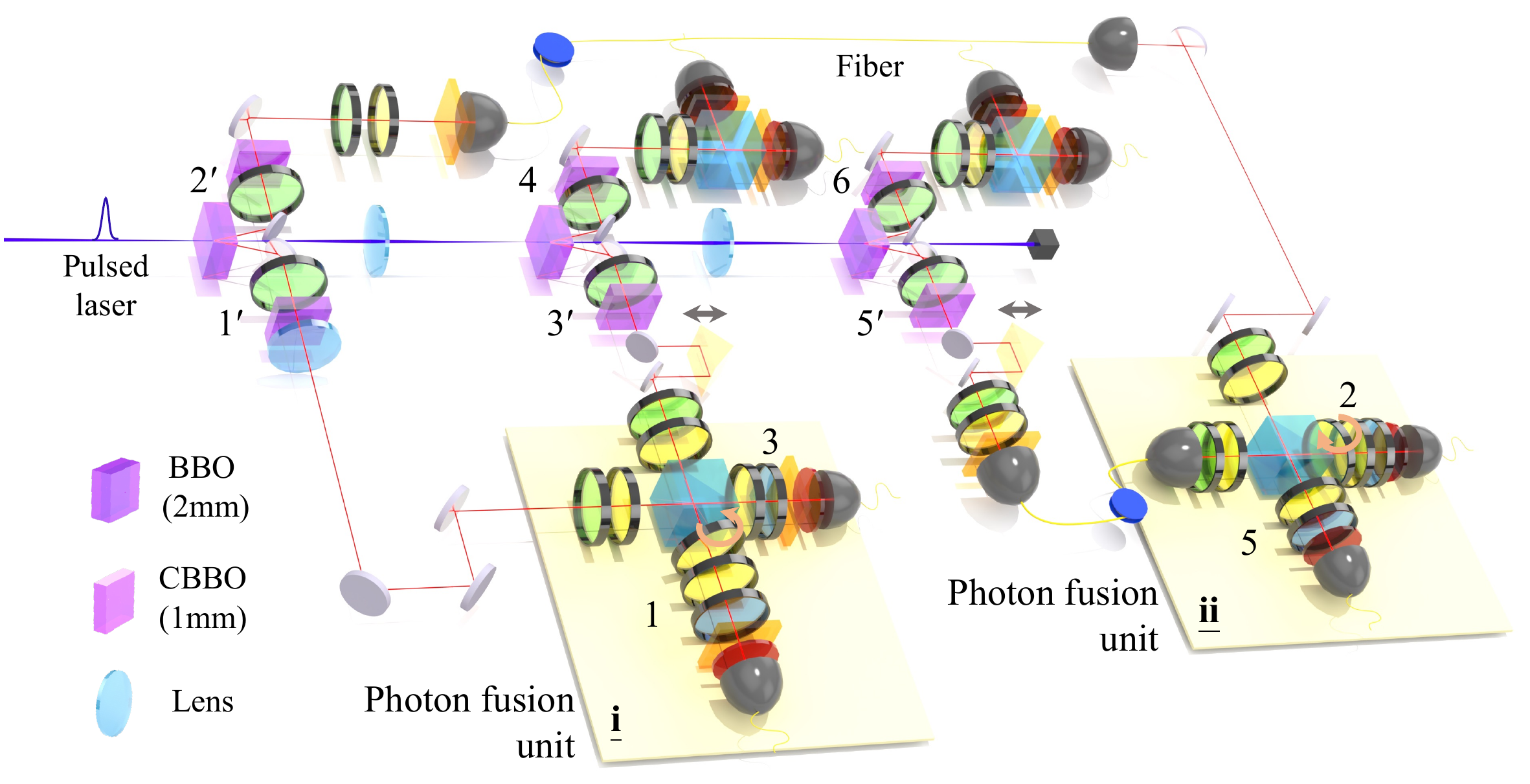}
\caption{Experimental setup of entangling six photons with two units of photon fusion. The three spontaneous parametric down-conversion (SPDC) sources~\cite{Kwiat95} were pumped with a pulsed laser beam (center wavelength 390 nm, pulse duration 144 fs, repetition rate 76 MHz, and average power 1.0 W). In each source, a $\beta$-barium-borate (BBO) crystal was pumped to generate pairs of polarization-entangled photons by type-II non-collinear SPDC, including the compensation for the walk-off effect by an additional correction BBO (CBBO) crystal. The prepared three photon pairs in modes $1'$-$2'$, $3'$-4, and $5'$-6 have an average creation rate of $\sim 86\times10^3$ pairs per second and an average fidelity of $(92.99\pm0.10)\%$ close to $\left|\psi^-\right\rangle=(1/\sqrt{2})(\left|HV\right\rangle-\left|VH\right\rangle)$, where $\left|H\right\rangle$ ($\left|V\right\rangle$) denotes the horizontal (vertical) photon polarization state. Two photon fusion units \textbf{i} and \textbf{ii} were set up as described in Fig.~\ref{Fig1}. The observed visibility of the Hong-Ou-Mandel type interference fringes~\cite{Hong87} is $\sim71\%$ ($\sim68\%$) measured in the diagonal$ (+/-)$ basis for the unit \textbf{i} (\textbf{ii}). The experimental interference imperfections are mainly caused by the high-order photon pair emission, spectrum mismatch between signal and idler photons, and the dispersion effect of photons in optical fiber transmission, which leads to pulse broadening.~(see Appendix~\ref{homimf}) All the experimental settings are tuned and aligned according to the target six-photon Greenberger-Horne-Zeilinger (GHZ) type entangled state~\cite{GHZ} $\left|G_6\right\rangle=(1/\sqrt{2})(\left|H_1V_2H_3V_4V_5H_6\right\rangle+\left|V_1H_2V_3H_4H_5V_6\right\rangle)$ in the present experiment. See Appendix~\ref{eps} for the detailed descriptions of the SPDC photon sources.}
\label{Fig2}
\end{figure*}

The existing method typically used for examining a photon fusion process in experiments relies on measuring the visibility of the desired two-photon coincidence-counting events conditioned on a given specific input photon pair state~\cite{Lu07,Pan12,Wang16,Zhong18,Adcock19,Llewellyn20,Wang20,MeyerScott22}. This examination is helpful but only provides partial information about the indistinguishability experimental interferometry can achieve rather than complete details of the fusion process for all possible input states. Recently, quantum process tomography~\cite{Chuang97,NielsenChuang} has been used to characterize the dynamics of experimental photon fusion completely~\cite{Bell12}. From a task-oriented viewpoint, knowing how to use such complete tomographic information to quantitatively assess the corresponding full entanglement creation ability of an entire experimental process thoroughly is essential. It also determines how to evaluate the related critical applications where photon fusion plays a role~\cite{Lu07,Pan12,Wang16,Zhong18,Adcock19,Llewellyn20,Wang20,MeyerScott22,Pan12,Wang16,Zhong18,Pan00,Lu20,Wu22,Walther05,Lu09,Yao12}. However, this still needs to be shown in the existing experimental tomographic photon fusion, where only partial process information based on target output entanglement under specific input photon states was considered~\cite{Bell12}. 

On the other hand, identifying elementary dynamical processes like fusing photon pairs is one of the fundamental goals uniting work on quantum information theory~\cite{NielsenChuang}. While significant and demanded, empirical evidence that the entire experimental photon fusion process can be quantified according to its abilities, such as creating entanglement, still needs to be provided. The theory of quantum process capability (QPC)~\cite{Hsieh17,Kuo19} describes using quantum process tomography for quantifying the full entanglement creation ability of experimental processes. This theory systematically shows that the experimentally obtained complete knowledge of photon fusion in terms of process matrix~\cite{Chuang97,NielsenChuang} can be quantified by sensible capability measures. In addition to the capability of entanglement creation, other process capabilities, such as entanglement preservation capability~\cite{Chen21}, can be quantified using the QPC theory. 

Here, we theoretically investigate and experimentally demonstrate the quantification of the entanglement creation capability and the entanglement preservation capability of fusing photon pairs. In addition to experiments following the standard QPC theory, we reveal for the first time that these process capabilities can be efficiently estimated without performing the total tomographic measurements. To illustrate the significance of quantifying photon fusion's capabilities, we show how it rigorously determines the generation of genuine multiphoton entanglement and EPR steering. This demonstration is adaptive to all photon-fusion-based applications reviewed at the beginning.

In the following, we will review the essential operation of photon fusion in Sec.~\ref{photon_fusion} and the QPC formalism in Sec.~\ref{QPC}. We then introduce a new theory for faithfully estimating the QPCs of photon fusion process in an experimentally efficient manner in Sec.~\ref{meqpc}. The experimental quantification of two photon fusion units is illustrated in Sec.~\ref{exp} afterward. With the measured photon fusion quantification, we show its important role in generating truly multiphoton entanglement and EPR steering in Sec.~\ref{Quantification}. The conclusion and outlook of the introduced concept and methods are given in Sec.~\ref{co}. Moreover, the technical details of the theoretical and experimental results are detailed in Appendices~\ref{ipd}-\ref{wspe}.

\section{Photon Fusion}\label{photon_fusion}
The fusion of photon pairs first coherently interferes with two individual photons in two different spatial modes ($a$ and $b$) at a PBS and then post-selects the two outputs in different modes ($c$ and $d$), as shown in the inset of Fig.~\ref{Fig1}. The PBS transmits horizontal ($H$) polarization and reflects vertical ($V$) polarization. That is, after the PBS, the two-photon states $\ket{H_aH_b},$ $\ket{H_aV_b},$ $\ket{V_aH_b},$ $\ket{V_aV_b}$ become $\ket{H_dH_c},$ $\ket{H_dV_d},$ $\ket{V_cH_c},$ $\ket{V_cV_d}$, respectively, where ${H_i}$ (${V_i}$) denotes $H$ ($V$) polarization in the spatial mode $i=a,$ $b,$ $c,$ $d$. A subsequent post-selection in the modes, $c$ and $d$, picks the states $\ket{H_cH_d}$ and $\ket{V_cV_d}$ only, which makes the fusion process of the photon pairs non-trace preserving.

Since the photon fusion superposes the two-photon amplitudes of different yet indistinguishable alternatives of the joint photodetection event in the different output modes of PBS, a quantum operation can represent an ideal photon fusion as described below~\cite{Lu07,Pan12,Wang16,Zhong18,Bell12}:
\begin{equation}\label{ideal_fusion}
\tilde{\rho}_\text{out}=\tilde{\chi}_\text{fusion}(\rho_\text{in})=M\rho_\text{in}M^\dagger,
\end{equation}
where $\rho_\text{in}$ and $\tilde{\rho}_\text{out}$ denote the input and output states of the photon fusion $\tilde{\chi}_\text{fusion}$, respectively, and
\begin{equation}\label{M}
M=\ket{H_{d}H_c}\!\!\bra{H_aH_b}+\ket{V_cV_d}\!\!\bra{V_aV_b},
\end{equation}
is called the fusion operator. Here $\tilde{\rho}_\text{out}$ is unnormalized, and its normalization factor ${\rm tr}(\tilde{\rho}_\text{out})$ is the probability of joint photodetection event. Hereafter, the symbol of $\tilde{\chi}_\text{fusion}$ also represents a un-normalized process matrix that completely describes the quantum operation of photon fusion. 

Notably, $\tilde{\chi}_\text{fusion}$ is \emph{nonclassical} because it is capable of creating and preserving entangled photons~\cite{Kuo19,Chen21}. For example, a separable input state of $\rho_\text{in}=\ketbra{\pm_a\pm_b}{\pm_a\pm_b}$, where $\ket{\pm_k}=(1/\sqrt{2})(\ket{H_k}\pm\ket{V_k})$ for $k=a,b$, becomes an entangled output state $\ket{\phi^{+}_\text{out}}$ with a probability of successful joint photodetection, $1/2$, i.e.,
\begin{equation}
\tilde{\chi}_\text{fusion}(\ket{\pm_a\pm_b}\!\!{\bra{\pm_a\pm_b}})=\frac{1}{2}\ketbra{\phi^{+}_{\text{out}}}{\phi^{+}_{\text{out}}},\nonumber
\end{equation} where $\ket{\phi^{+}_\text{out}}=(1/\sqrt{2})(\ket{H_cH_d}+\ket{V_cV_d})$. Moreover, let $\rho_\text{in}$ be an entangled state of $\ketbra{\phi^{+}_{\text{in}}}{\phi^{+}_{\text{in}}}$, where $\ket{\phi^{+}_{\text{in}}}=(1/\sqrt{2})(\ket{H_aH_b}+\ket{V_aV_b})$, the output state is entangled as well: $\tilde{\chi}_\text{fusion}(\ketbra{\phi^{+}_{\text{in}}}{\phi^{+}_{\text{in}}})=\ketbra{\phi^{+}_{\text{out}}}{\phi^{+}_{\text{out}}}$.

\section{Capabilities of photon fusion}\label{QPC}

\subsection{Quantum process capabilities}

The theory proposed in Refs.~\cite{Kuo19,Chen21} predicts that the nonclassical characteristics of $\tilde{\chi}_\text{fusion}$ qualitatively described above can be quantified further as QPCs of entanglement creation and preservation. Such QPCs can unambiguously determine precisely the extent to which an imperfect experimental photon fusion, denoted as $\tilde{\chi}_\text{expt}$, can create and preserve entangled photons. According to the theory~\cite{Kuo19,Chen21}, $\tilde{\chi}_\text{expt}$ is quantified by comparing it with processes without the prescribed quantum capabilities, called incapable processes and denoted by $\chi_\mathcal{I}$. 

The following two different sensible measures quantify photon fusion by:
(a) Composition. 
\begin{equation}
\chi_{\text{expt}} = \alpha\chi_{\mathcal{C}} + (1 - \alpha)\chi_{\mathcal{I}},\label{c}
\end{equation}
where $\alpha=\alpha_{\text{cre}}$ $(\alpha_{\text{pre}})$ represents the minimum amount of the process with the prescribed quantum capability of entanglement creation (preservation), denoted as $\chi_\mathcal{C}$, that can be found in $\chi_{\text{expt}}=\tilde{\chi}_\text{expt}/{\rm tr}(\tilde{\chi}_\text{expt})$. As $\alpha=0$, $\chi_{\text{expt}}$ is considered as incapable of creating (preserving) entangled photon pair. Whereas $\alpha>0$ implies capable $\chi_{\text{expt}}$, and $\alpha=1$ showing the maximum QPC of entanglement creation (preservation) that $\chi_{\text{expt}}$ has. (b) Robustness.
\begin{equation}
\dfrac{\chi_{\text{expt}} + \beta\chi'}{1 + \beta} = \chi_{\mathcal{I}},\label{betab}
\end{equation}
where $\beta=\beta_{\text{cre}}$ ($\beta_{\text{pre}}$) is the minimum amount of noise process, $\chi'$, (positive operator) added such that $\chi_{\text{expt}}$  becomes incapable of creasing (preserving) entanglement. As $\beta=0$, $\chi_{\text{expt}}$ is identified as incapable. Then $\beta>0$ describes a capable process with the prescribed entanglement creation (preservation) quantum capability. 

In addition to the above two QPC measures, $\chi_{\text{expt}}$ can be examined by the following process fidelity criterion that
\begin{equation}
F_\text{expt}={\rm tr}(\chi_{\text{expt}}\chi_{\text{fusion}})>F_{\mathcal{I}}=\max_{\chi_{\mathcal{I}}}[{\rm tr}(\chi_{\mathcal{I}}\chi_{\text{fusion}})],
\end{equation}
indicating that $\chi_{\text{expt}}$ is a capable process close to the target process $\chi_{\text{fusion}}$.

When $\chi_{\text{expt}}$ is given, the quantities of $\alpha$ and $\beta$ can be determined by using semidefinite programming (SDP) to solve the corresponding optimization problems under a set of constraints, denoted as $\mathcal{D}_{\mathcal{I}}$, which details the incapable processes $\chi_{\mathcal{I}}$, and the ones for $\chi_{\text{expt}}$ and $\chi'$. In obtaining $\alpha$ and $\beta$ via SDP, the choice of $\chi_{\mathcal{I}}$ is the optimization parameters. With the SDP solver, the specific incapable process matrix $\chi_{\mathcal{I}}$ that achieves the optimal value $\alpha$ and $\beta$ respectively in Eq.~(\ref{c}) and Eq.~(\ref{betab}) can be determined. The threshold $F_{\mathcal{I}}$ can be calculated in the same manner as $\alpha$ and $\beta$. See Appendix~\ref{ipd} for the complete descriptions of the constraint sets $\mathcal{D}_{\mathcal{I}}$ and their use in measuring QPCs detailed in Appendix~\ref{mqpc}.

\begin{figure*}[t]
\includegraphics[width=16cm]{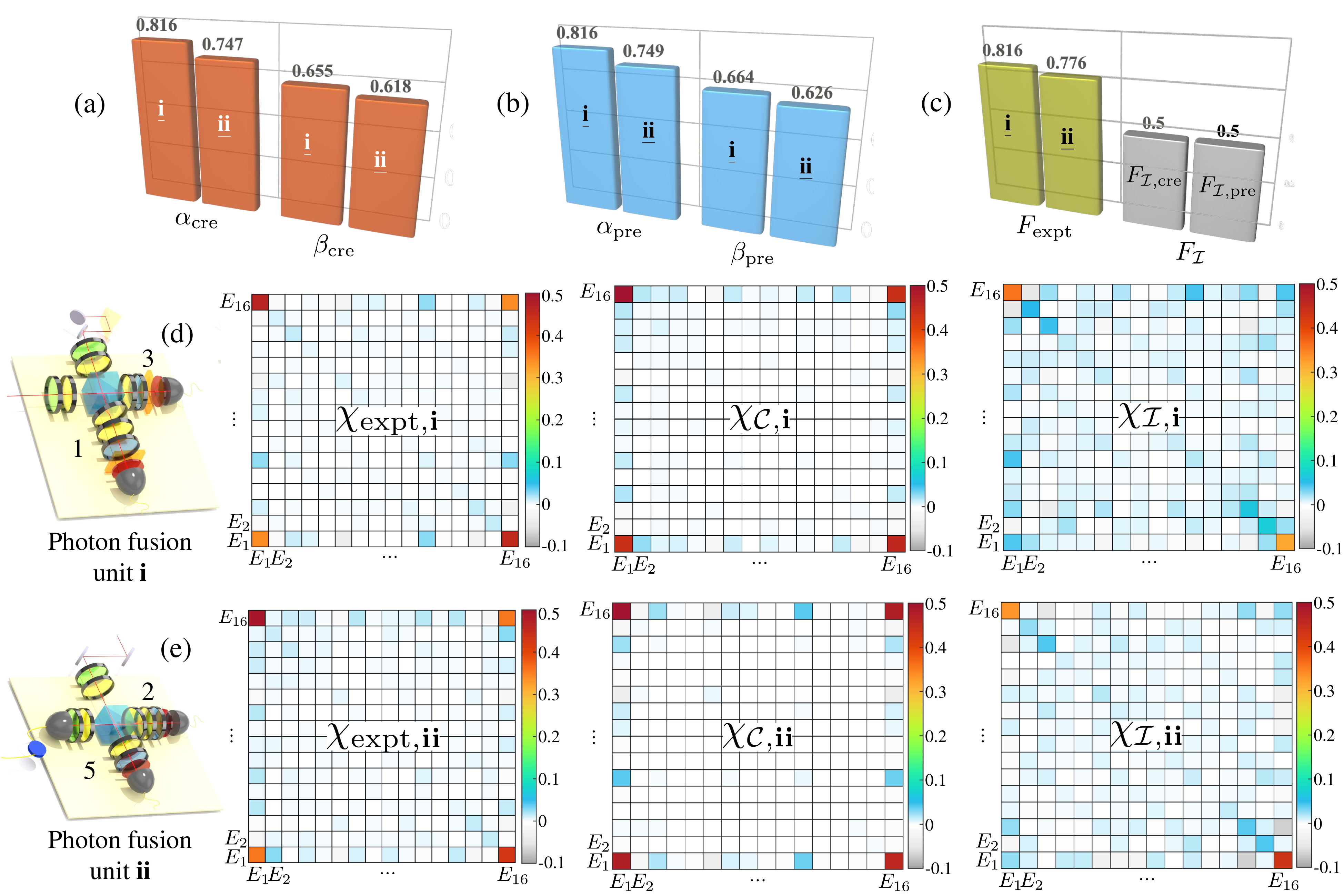}
\caption{The experimental photon fusion capabilities for six-photon entanglement. (a) The entanglement creation capabilities for the experimental photon fusion units \textbf{i} and \textbf{ii} are quantified by the composition, $\alpha_{\text{cre}}$, and robustness, $\beta_{\text{cre}}$. (b) The experimental entanglement preservation capabilities for \textbf{i} and \textbf{ii} are quantified by the composition $\alpha_{\text{pre}}$ and the robustness $\beta_{\text{pre}}$. By (a) and (b) the portion that cannot generate entanglement but can preserve entanglement can be quantified further by $\alpha_{\rm pre'} = \alpha_{\rm pre}-\alpha_{\rm cre}$~(see Appendix~\ref{mqpc}). (c) The measured values of the process fidelities for both units are better than the maximum values any incapable process of entanglement creation ($F_{\mathcal{I},\text{cre}}$) and entanglement preservation ($F_{\mathcal{I},\text{pre}}$) can achieve, $F_{\mathcal{I}}$. (a)-(c) are obtained using the QPC theory to thoroughly quantify the measured process matrices $\chi_{\text{expt,\textbf{i}}}$ and $\chi_{\text{expt,\textbf{ii}}}$ for the experimental photon fusion units \textbf{i} and \textbf{ii}, respectively. (d) and (e) illustrate the real parts of their process matrices. See Appendix~\ref{mpm} for their theoretical and experimental details of measurements of the process matrices and their maximum likelihood estimation \cite{James01,OBrien04}. The real parts of $\chi_{\mathcal{C},k}$ and $\chi_{\mathcal{I},k}$, for $k=\textbf{i}, \textbf{ii}$, respectively corresponding to capable and incapable processes that are obtained via SDP solver and achieve $\alpha_{\rm cre}$ of $\chi_{\text{expt,}k}$ are illustrated. See~Eq.~(\ref{fa}). Since the matrix elements in the imaginary parts of all the process matrices, $\chi_{\text{expt,\textbf{i}}},\chi_{\text{expt,\textbf{ii}}},\chi_{\mathcal{C},k},\chi_{\mathcal{I},k}$, are small, all the imaginary parts are neglected here.}
\label{Fig3}
\end{figure*}

\section{Estimation of the QPC}\label{meqpc}

The QPC formalism is adaptive to estimating $\alpha$ and $\beta$ without using the full process tomography data. Experimentally, measuring QPCs demands quantum process tomography on $\tilde{\chi}_\text{expt}$, where $16$ specific states of $\rho_\text{in}$ are prepared and the corresponding outputs $\tilde{\rho}_\text{out}=\tilde{\chi}_\text{expt}(\rho_\text{in})$ are required 9 local measurements for tomography analysis. Since each local measurement considers 4 output events, it totally requires $16\cdot9\cdot4=576$ measurements of the probabilities of input-output-state events. Whereas, given a subset of these probabilities denoted by $\{\mathcal{P}\}$, it is possible to find an estimated process matrix, denoted by $\tilde{\chi}_\text{SDP}$, and its $\alpha$ and $\beta$ via SDP under $\mathcal{D}_{\mathcal{I}}$ and the constraints of the given probabilities $\{\mathcal{P}\}$.

A concrete example of $\{\mathcal{P}\}$ is the probabilities used for estimating the lower bound of process fidelity, $F_{\rm LB}$, where only $40$ probabilities are used. It is determined by: $F_{{\rm LB}}=(1/2)(F^{\rm expt}_{Z \rightarrow Z} +F^{\rm expt}_{X \rightarrow X} +F^{\rm expt}_{X \rightarrow Y} -1)$, where $F^{\rm expt}_{Z \rightarrow Z}$, $F^{\rm expt}_{X \rightarrow X}$, and $F^{\rm expt}_{X \rightarrow Y}$ are three classical fidelities~\cite{Hofmann05,Hofmann06}. See Appendix~\ref{eqpc} for the complete descriptions used in our experiments. As the errors conserve horizontal/vertical polarization, $F_{{\rm LB}}$ can be considered as the process fidelity of photon fusion~\cite{Okamoto08}.

With the method introduced above, one can estimate $\alpha$ and $\beta$ of the process $\tilde{\chi}_\text{SDP}$ given the probabilities for the process fidelity lower bound in a very efficient manner. See Appendix~\ref{eqpc} detailing how to estimate QPCs using the given probabilities and SDP.

Furthermore, in Sec.~\ref{Quantification}, we will illustrate that the QPC estimates are helpful to jointly evaluate whether the experimental photon fusion units are capable of creating genuine multiphoton quantum correlations before performing multiphoton measurements.

\section{Experimental quantification of photon fusion}\label{exp}

The experimental setup of a unit for photon fusion is shown in Fig.~\ref{Fig1}. As will be described below, we set up two units of photon fusion in our six-photon experiment. See the units \textbf{i} and \textbf{ii} in Fig.~\ref{Fig2}. To begin with, we examine the QPCs of the first unit of photon fusion (the unit \textbf{i}). The photon in the input mode $a=1'$ of the first entangled pair from spontaneous parametric down-conversion (SPDC) acts as a signal, and the photon in $2'$ is a trigger. Similarly, photon in the input mode $b=3'$ of the second SPDC entangled pair is ready as the photon $4$ is detected.

First, we measure the QPCs of the photon fusion under the best photon interference at the PBS. This condition was achieved by scanning the delay and tuning the QWP such that the visibility of the four-fold coincidence counts detected on the diagonal basis, $\{\ket{+},\ket{-}\}$, is maximum. By tomographically deconstructing the output states of specified input states, $\chi_{\text{expt}}$ was then measured (see Appendix~\ref{mpm}), and by which its QPCs were then determined, including the capabilities of entanglement creation and preservation. See Figs.~\ref{Fig3}(a)-(c). The second unit of photon fusion (the unit \textbf{ii}) is quantified in the same manner.

Specifically, we have demonstrated that, according to Eq.~(\ref{c}), each experimental photon fusion unit can be quantified by:
\begin{equation}
\chi_{\text{expt},k}=\alpha_{k}\chi_{\mathcal{C},k}+(1-\alpha_{k})\chi_{\mathcal{I},k},\label{fa}
\end{equation}
for $k=\textbf{i}, \textbf{ii}$, where $\alpha_{k}$ is the entanglement creation composition $\alpha_{\text{cre}}$ of the photon fusion unit $k$, and $\chi_{\mathcal{C},k}$ and $\chi_{\mathcal{I},k}$ are the corresponding capable and incapable processes, respectively. Figures~\ref{Fig3}(d) and \ref{Fig3}(e) illustrate the process matrices of $\chi_{\mathcal{C},k}$ and $\chi_{\mathcal{I},k}$ which constitute the two photon fusion units $\chi_{\text{expt},k}$ with respect to the entanglement-creation QPC. As will be shown in Sec.~\ref{Quantification}, composition $\alpha_{k}$ determines genuine multipartite quantum correlations of multiphoton interferometry outputs.  

\begin{figure}[t]
\includegraphics[width=8.5cm]{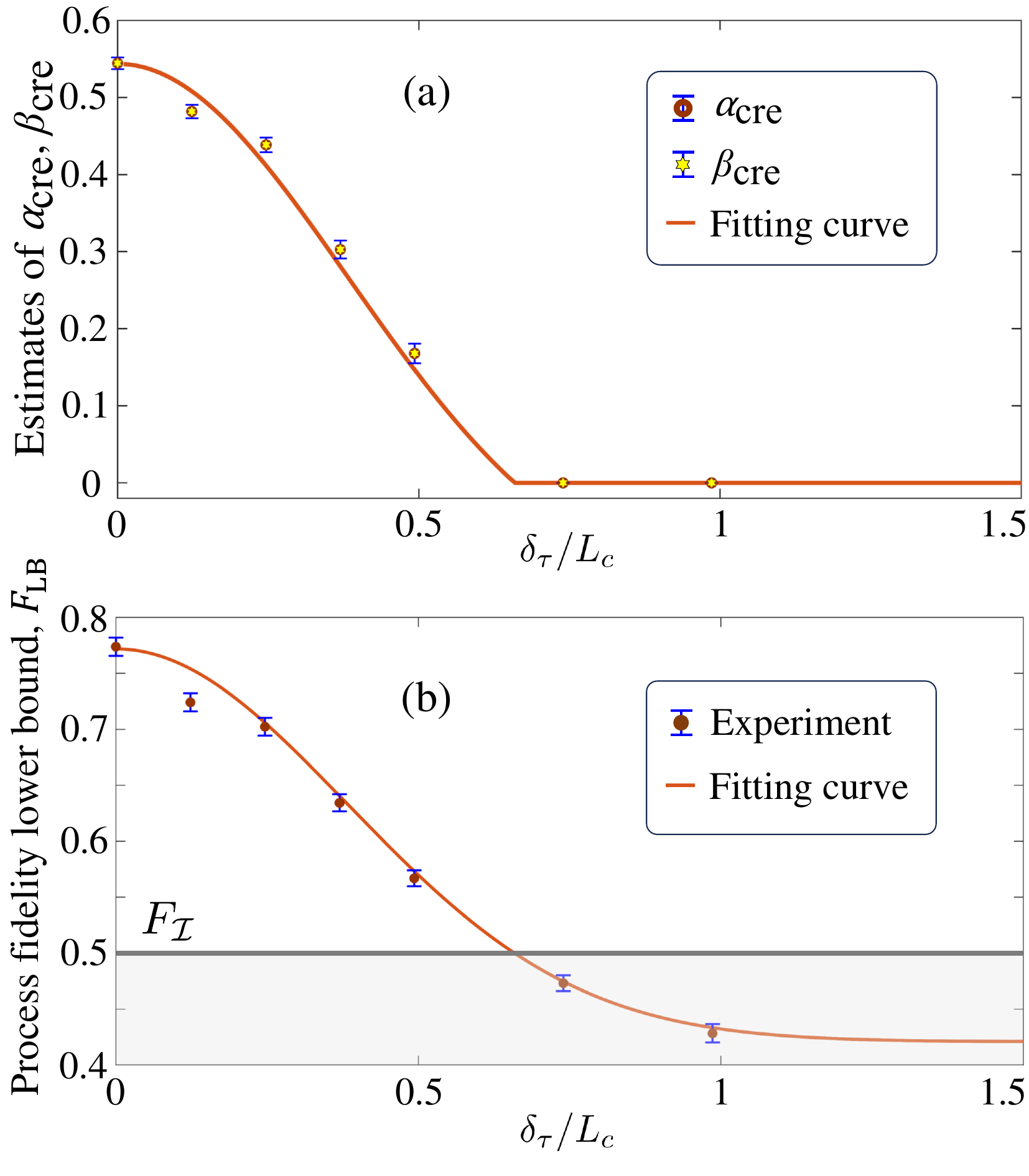}
\caption{Experimental transition between capable and incapable of entanglement creation processes. With the measured QPC estimates (a) and the process fidelity lower bound, $F_{\rm LB}$, (b) for the first photon fusion unit \textbf{i}, we show how the photon fusion's entanglement creation capability decreases with the increasing interferometry delay, $\delta_\tau$. The symbols representing $\alpha_{\text{cre}}$ and $\beta_{\text{cre}}$ overlap since we have $\alpha_{\text{cre}}=\beta_{\text{cre}}$ in the present case. As seen, the QPC measures $\alpha$ and $\beta$ in (a) reliably indicate that the experimental photon fusion becomes incapable as the delay is close to but not exceed one coherence length, consistent with the examination using the fidelity criterion (b). The used narrow-band filter (central wavelength $\lambda=780$ nm, transmission rate $>99\%$) with the bandwidth (full-width at half maximum) of $\lambda_{\text{FWHM}}=3$ nm determines the coherence length of $L_c\sim\lambda^2/\lambda_{\text{FWHM}}=203$ $\mu$m. Moreover, we calculate the uncertainty in $F_{{\rm LB}}$, denoted by $\Delta F_{{\rm LB}}$, by using the propagation of errors and the Poissonian error in the coincidence counts $N_i$ in $F_{{\rm LB}}$. That is, $\Delta F_{{\rm LB}}=\sqrt{\sum_{i}(\Delta N_i\frac{\partial F_{{\rm LB}}}{\partial N_i})^2}$, where $\Delta N_i=\sqrt{N_i}$. The lower bound with errors $F_{{\rm LB}}\pm\Delta F_{{\rm LB}}$ is used to obtain two $\chi_{\rm SDP}$ corresponding to $F_{{\rm LB}}+\Delta F_{{\rm LB}}$ and $F_{{\rm LB}}-\Delta F_{{\rm LB}}$, respectively. The $\alpha$ and $\beta$ obtained from these two estimated process matrices are considered errors of $\alpha$ and $\beta$, as depicted in (a).}
\label{Fig4}
\end{figure}

Furthermore, the newly introduced method to efficiently estimate QPC makes it more experimentally feasible to demonstrate the transitions between capable and incapable photon interferometry. We measured the estimated $\alpha$ and $\beta$ under three sets of classical fidelities consisting of $40$-probability set $\{\mathcal{P}\}$ in different delay values for capable and incapable processes. See Fig.~\ref{Fig4}. This also demonstrates that the measured entanglement-creation QPCs are faithful to revealing experimental imperfections. They monotonically decrease with the increase of delay.

It is worth noting that while the capability of $\alpha_{\text{cre}}\sim0.545$ for $\delta_\tau=0$ in Fig.~\ref{Fig4}(a) does not match the capability value $\alpha_{\text{cre}}\sim0.816$ [Fig.~\ref{Fig3}(a)] derived from the complete process matrix, this quantity reveals the capability that a process with a given fidgety lower bound can have through our method. In particular, measuring this capability corresponding to the process lower bound is much more efficient. Therefore, our introduced concept and methods provide the possibility of a trade-off between measurement efficiency and capability evaluation accuracy. With such a trade-off, one can acquire preliminary information about the capability using a few measurements without completing the whole process of tomography measurements. One can also utilize such preliminary information to evaluate the capability of the photon fusions for generating genuine multiphoton quantum correlations. See a concrete illustration in Sec.~\ref{six} regarding Eq.~(\ref{assessix}).

\section{QPC for genuine multiphoton quantum correlations}\label{Quantification}

\subsection{Genuine multiphoton entanglement \\and EPR steering}

Since photon fusion is intrinsically necessary for creating multiphoton entangled states, QPC provides a quantitatively precise method to assess multiphoton entanglement generation in photon-fusion-based experiments, including correlation creation of genuine multipartite EPR steering~\cite{He13,Li15,Armstrong15,Cavalcanti15}. Suppose $n$ units of photon fusion are set up in an interferometry on $n+1$ pairs of entangled photons. With Eq.~(\ref{c}), the output states can be represented as
\begin{equation}
\tilde{\rho}_{\rm out}=\prod_{k=1}^{n}\alpha_{{\rm cre},k}\chi_{\mathcal{C},k}(\rho_\text{in})+\tilde{\rho}_{\text{sep}},\label{gme}
\end{equation}
where $\alpha_{{\rm cre},k}$ is the $k$th photon fusion unit's entanglement creation composition, and $\tilde{\rho}_{\text{sep}}$ is an unnormalized bi-separable state. Here, the product of composition: $\prod_{k=1}^{n}\alpha_{{\rm cre},k}$, determines the genuine $N$-photon entanglement composition in $\tilde{\rho}_{\rm out}$ for even $N=2(n+1)$. (Equation~(\ref{gme}) and other conclusions given below are applicable to the cases of odd $N$, where one of the $n+1$ pairs is replaced by a single photon state.) To investigate this conclusion for the output state $\tilde{\rho}_{\rm out}$ in Eq.~(\ref{gme}), let us first consider our experimental photon fusion processes described by Eq.~(\ref{fa}) as a concrete illustration.

The two photon fusion units' capability compositions determine the component of the created genuine six-photon multipartite entanglement in the created photons. That is, after operation of the two photon fusion units, the input of three photon pairs in the experiment becomes
\begin{eqnarray}
&&\tilde{\rho}_{\rm out}=\chi_{\text{expt,\textbf{ii}}}\chi_{\text{expt,\textbf{i}}}(\rho_{1'2'}\otimes\rho_{3'4}\otimes\rho_{5'6})\nonumber\\
&&\ \ \ \ \ \ =\alpha_{\textbf{i}}\alpha_{\textbf{ii}}\chi_{\mathcal{C},\textbf{ii}}\chi_{\mathcal{C},\textbf{i}}(\rho_{1'2'}\!\otimes\!\rho_{3'4}\!\otimes\!\rho_{5'6})\!+\!\tilde{\rho}_{\text{sep}}.\label{sixoutput}
\end{eqnarray}
Since $\chi_{\mathcal{C},\textbf{i}}$ and $\chi_{\mathcal{C},\textbf{ii}}$ are capable of creating entanglement, the state $\tilde{\rho}_{\mathcal{C}}=\chi_{\mathcal{C},\textbf{ii}}\chi_{\mathcal{C},\textbf{i}}(\rho_{1'2'}\otimes\rho_{3'4}\otimes\rho_{5'6})$ is genuinely six-partite entangled as the three-pair initial state is properly prepared, which describes the only multipartite entanglement component in the resulting output state $\tilde{\rho}_{\rm out}$. Therefore, the compositions $\alpha_{\textbf{i}}$ and $\alpha_{\textbf{ii}}$ by $\alpha_{\textbf{i}}\alpha_{\textbf{ii}}$ together quantify the interferometry performance for creating genuinely six-partite entangled photons. Compared to the existing methods based on process fidelity~\cite{Bell12,Hofmann05,Hofmann06,Okamoto08}, our presented method provides quantitatively precise interferometry information.

Also, according to Eq.~(\ref{sixoutput}) and the characteristics of $\chi_{\mathcal{C},\textbf{i}}$ and $\chi_{\mathcal{C},\textbf{ii}}$ discussed above, $\alpha_{\textbf{i}}$ and $\alpha_{\textbf{ii}}$ decide the similarity between the created photons and a target genuinely multiphoton entangled state in terms of the state fidelity, defined by $F_{s,N}=\bra{G_N}\rho_{G_N}\ket{G_N}$, where $\rho_{G_N}$ for $N=6$ in the present case is a normalized state of $\tilde{\rho}_{\rm out}$. For example, if either $\alpha_{\textbf{i}}=0$ or $\alpha_{\textbf{ii}}=0$, the output state is bi-separable and the resulting best state fidelity is $F_{s}^{({\rm sep})}=1/2$~\cite{Bourennane04}. A fidelity better than this threshold means creating genuine multipartite entangled states. Thus, $\alpha_{\textbf{i}}$ and $\alpha_{\textbf{ii}}$ should be high enough for both units of photon fusion when using the entanglement witness condition $F_{s,N}>1/2$ as the criterion for creating genuine six-photon entanglement close to the GHZ state.

As shown in Refs.~\cite{Lu20,Kao23}, pre-existing classical data from classical networking nodes or uncharacterized measurement devices can cause the resulting fidelity to be unreliable even for $F_{s,N}>1/2$. This makes false positives of entanglement-witness-based entanglement detection in quantum networks. To rule out the mimicry of the strategy using pre-existing classical data, it has been shown that the state fidelity should satisfy $F_{s,N}>F_{s}^{(\rm pre)}=(1/4)(1+\sqrt{3})\approx0.683$ for even $N$, regardless of state tomography measurements either on the Pauli observables~\cite{Lu20} or on the $N+1$ minimum number of observables~\cite{Kao23}. As satisfied by the experimental results, the created state possesses genuine $N$-photon EPR steerability. Then, this EPR steering criterion on $\alpha_{\textbf{i}}$ and $\alpha_{\textbf{ii}}$ is stricter than the entanglement witness.

\subsection{Composition criteria for genuine multiphoton quantum correlations}

According to the above discussions about Eq.~(\ref{sixoutput}) and the state-fidelity criterion for genuine $N$-photon entanglement and EPR steering, we, therefore, arrive at the following condition on $n$-photon-fusion process for creating genuine $N$-photon entanglement and EPR steering in terms of the compositions:
\begin{equation}
\prod_{k=1}^{n}\alpha_{{\rm cre},k}>\frac{{\rm tr}(\tilde{\rho}_{\rm out})(\mathcal{F}_s-F_{s,{\rm sep}})}{{\rm tr}(\tilde{\rho}_{\mathcal{C}})(F_{s,{\rm \mathcal{C}}}-F_{s,{\rm sep}})},\label{acriterion}
\end{equation}
where $\tilde{\rho}_{\mathcal{C}}=\prod_{k=1}^{n}\chi_{\mathcal{C},k}(\rho_\text{in})$, and $F_{s,{\rm \mathcal{C}}}$ and $F_{s,{\rm sep}}$ denote the state fidelities of the target $N$-qubit GHZ-type state $\ket{G_N}$ and the normalized state of $\tilde{\rho}_{\mathcal{C}}$ and $\tilde{\rho}_{\text{sep}}$, respectively. Equation~(\ref{acriterion}) is derived from the condition: $\bra{G_N}\rho_{\rm out}\ket{G_N}>\mathcal{F}_s$, where $\rho_{\rm out}$ is the normalized state of $\tilde{\rho}_{\rm out}$ defined in Eqs.~(\ref{gme}),(\ref{sixoutput}). If an experimental result of composition product satisfies the criterion~(\ref{acriterion}) for $\mathcal{F}_s=F_{s}^{({\rm sep})}$, then $\rho_{\rm out}$ is genuinely $N$-photon entangled. As the $n$ photon fusions satisfy the criterion further for $\mathcal{F}_s=F_{s}^{({\rm pre})}$, then the photon fusion units enable a joint creation of states which can show genuine $N$-photon EPR steering effect. In what follows, we will demonstrate how Eq.~(\ref{acriterion}) is helpful to examine the experimental photon fusion units.

\subsubsection{Genuine six-photon entanglement}\label{six}

As the three created photon pairs, $\alpha_{\textbf{i}}$, and $\alpha_{\textbf{ii}}$ are given, including the processes $\chi_{\mathcal{C},k}$ and $\chi_{\mathcal{I},k}$ for $k=\textbf{i}, \textbf{ii}$ in Eq.~(\ref{fa}), the above criterion helps evaluate the experimental photon fusion units before performing six-photon measurements. In our experiment, the created entangled photon pairs are close to the Werner-type state with a type-II down-conversion dephasing noise. With this condition for the three pair states: $\rho_{1'2'}$, $\rho_{3'4}$, $\rho_{5'6}$, and the determined $\alpha_{\textbf{i}}$ and $\alpha_{\textbf{ii}}$ [Fig.~\ref{Fig3}(a)], we use Eq.~(\ref{acriterion}) to examine the two-photon fusion units and find that they are qualified for generating six-photon entanglement by:
\begin{equation}
\alpha_{\textbf{i}}\alpha_{\textbf{ii}}\approx0.610>\frac{{\rm tr}(\tilde{\rho}_{\rm out})(0.5-F_{s,{\rm sep}})}{{\rm tr}(\tilde{\rho}_{\mathcal{C}})(F_{s,{\rm \mathcal{C}}}-F_{s,{\rm sep}})}\approx0.446.\label{asssix}
\end{equation}
 where ${\rm tr}(\tilde{\rho}_{\rm out})=0.062$, ${\rm tr}(\tilde{\rho}_{\mathcal{C}})=0.062$, $F_{s,{\rm sep}}=0.288$, and $F_{s,{\rm \mathcal{C}}}=0.763$.
 
 Alternatively, this means that the state fidelity estimated using Eq.~(\ref{sixoutput}) is: $F_{s,6}\sim0.58$, which satisfies the fidelity criterion. See Appendix~\ref{spsf} for the detailed derivation. The resulting state is, therefore, identified as genuinely six-photon entangled. This is consistent with the experimental result examined by measuring a two-measurement entanglement witness~\cite{Toth05} in our experiment. See Appendix~\ref{wspe} for the detailed experimental entanglement witness results (Fig.~\ref{witness6photon}). However, the photon fusion units' joint performance does not satisfy the criterion for showing genuine six-photon EPR steering.
 
 It is worth noting that, when assessing the estimates of $\alpha_{\rm cre}$ introduced in Sec.~\ref{meqpc} via Eq.~(\ref{acriterion}), $\alpha_{\textbf{i}}$ and $\alpha_{\textbf{ii}}$ are also identified as being qualified for the case of zero delay ($\delta_{\tau}=0$) by:
 \begin{equation}
\alpha_{\textbf{i}}\alpha_{\textbf{ii}}\approx0.285>\frac{{\rm tr}(\tilde{\rho}_{\rm out})(0.5-F_{s,{\rm sep}})}{{\rm tr}(\tilde{\rho}_{\mathcal{C}})(F_{s,{\rm \mathcal{C}}}-F_{s,{\rm sep}})}\approx0.245,\label{assessix}
\end{equation}
where $\alpha_{\textbf{i}}=0.545$, $\alpha_{\textbf{ii}}=0.524$, ${\rm tr}(\tilde{\rho}_{\rm out})=0.062$, ${\rm tr}(\tilde{\rho}_{\mathcal{C}})=0.062$, $F_{s,{\rm sep}}=0.379$, and $F_{s,{\rm \mathcal{C}}}=0.872$. Therefore, using only three sets of classical fidelities consisting of $40$-probability set $\{\mathcal{P}\}$ is sufficient to infer from Eq.~(\ref{assessix}) that the photon fusion units jointly enables the multiphoton interferometry to generate truly six-photon entangled GHZ type state. This is also consistent with the result of Eq.~(\ref{asssix}) based on using full process tomography to obtain $\alpha_{\textbf{i}}$ and $\alpha_{\textbf{ii}}$.
 
 \subsubsection{Genuine four-photon EPR steering}
 
 Since satisfying the criterion for showing genuine multipartite EPR steerability requires higher performance of $n$ joint photon fusion units~(\ref{acriterion}), it becomes difficult to observe such high-order EPR steering effect with increasing $N$. For $n=1$ and $N=4$, we successfully demonstrate genuine four-photon EPR steering by using the first photon fusion unit and the pair states: $\rho_{1'2'}$ and $\rho_{3'4}$, and by using the second photon fusion unit and the pair states: $\rho_{1'2'}$ and $\rho_{5'6}$. Their measured state fidelities are: $(75.79\pm0.37)\%$ and $(71.91\pm0.49)\%$, respectively, by measuring the minimum five observables~\cite{Kao23}. They are consistent with the assessments based on the criterion~(\ref{acriterion}) by:
 \begin{equation}
\alpha_{\textbf{i}}\approx0.816>\frac{{\rm tr}(\tilde{\rho}_{\rm out})(0.683-F_{s,{\rm sep}})}{{\rm tr}(\tilde{\rho}_{\mathcal{C}})(F_{s,{\rm \mathcal{C}}}-F_{s,{\rm sep}})}\approx0.746,\label{0.816}
\end{equation} 
where ${\rm tr}(\tilde{\rho}_{\rm out})=0.249$, ${\rm tr}(\tilde{\rho}_{\mathcal{C}})=0.249$, $F_{s,{\rm sep}}=0.345$, and $F_{s,{\rm \mathcal{C}}}=0.798$, and
\begin{equation}
\alpha_{\textbf{ii}}\approx0.747>\frac{{\rm tr}(\tilde{\rho}_{\rm out})(0.683-F_{s,{\rm sep}})}{{\rm tr}(\tilde{\rho}_{\mathcal{C}})(F_{s,{\rm \mathcal{C}}}-F_{s,{\rm sep}})}\approx0.712,\label{0.747}
\end{equation}
where ${\rm tr}(\tilde{\rho}_{\rm out})=0.249$, ${\rm tr}(\tilde{\rho}_{\mathcal{C}})=0.249$, $F_{s,{\rm sep}}=0.331$, and $F_{s,{\rm \mathcal{C}}}=0.825$. Here, we use the same method as was employed with the six-photon case~(\ref{asssix}).

Moreover, we have used the criterion~(\ref{acriterion}) to evaluate the estimates of $\alpha_{\textbf{i}}$ and $\alpha_{\textbf{ii}}$ for genuine multipartite EPR steering. The first photon fusion unit is capable of creating genuine four-photon steering and identified by:
 \begin{equation}
\alpha_{\textbf{i}}\approx0.545>\frac{{\rm tr}(\tilde{\rho}_{\rm out})(0.683-F_{s,{\rm sep}})}{{\rm tr}(\tilde{\rho}_{\mathcal{C}})(F_{s,{\rm \mathcal{C}}}-F_{s,{\rm sep}})}\approx 0.526,
\end{equation} 
where ${\rm tr}(\tilde{\rho}_{\rm out})=0.249$, ${\rm tr}(\tilde{\rho}_{\mathcal{C}})=0.249$, $F_{s,{\rm sep}}=0.452$, and $F_{s,{\rm \mathcal{C}}}=0.891$. This is consistent with the identification results of Eq.~(\ref{0.816}). However, the $40$-probability set $\{\mathcal{P}\}$ is still insufficient to support showing that the second unit is capable by the following result:
\begin{equation}
\alpha_{\textbf{ii}}\approx0.524<\frac{{\rm tr}(\tilde{\rho}_{\rm out})(0.683-F_{s,{\rm sep}})}{{\rm tr}(\tilde{\rho}_{\mathcal{C}})(F_{s,{\rm \mathcal{C}}}-F_{s,{\rm sep}})}\approx 0.542,
\end{equation}
where ${\rm tr}(\tilde{\rho}_{\rm out})=0.249$, ${\rm tr}(\tilde{\rho}_{\mathcal{C}})=0.249$, $F_{s,{\rm sep}}=0.450$, and $F_{s,{\rm \mathcal{C}}}=0.880$. For this case, we need the complete analysis for $\alpha_{\textbf{ii}}$ as shown in the identification of Eq.~(\ref{0.747}).

\section{Conclusion and outlook}~\label{co}

Photon fusion is the kernel of creating multiphoton entanglement for various quantum foundations and quantum information applications. We theoretically investigate and experimentally demonstrate the quantification of photon fusion's capabilities of creating and preserving entangled photon pairs regarding capability composition, robustness, and fidelity criterion. We also show that these quantum capabilities can be efficiently estimated to faithfully reveal the interferometry imperfections in the experiment for the first time. Furthermore, we use capability quantification to assess the experimental multiphoton interferometry (i.e., the joint performance of all the photon fusion units) quantitatively in generating genuine six-photon entanglement. We prove that the amount of genuine multipartite entanglement and EPR steering in the created photons and state fidelity is determined by the capability composition of the fusion processes.

Since these results are implementable using photonic tomography, the concept and methods can be utilized for benchmarking photon fusion in the existing quantum photonics engineering where photon fusion served as a building block~\cite{Lu07,Pan12,Wang16,Zhong18,Adcock19,Llewellyn20,Wang20,MeyerScott22,Pan12,Wang16,Zhong18,Pan00,Lu20,Wu22,Walther05,Lu09,Yao12}. Moreover, genuine multipartite entanglement powers quantum networks' correlation, coordination, and security~\cite{Kimble08,Ritter12,Wehner18}. For practical implementation, it may be helpful to use these findings further to investigate the relationship between photon fusions' capabilities and the criteria for achieving multiphoton interferometry under untrusted experimental apparatuses~\cite{Huang23} performed in uncharacterized network nodes~\cite{Lu20,Kao23}.

\section*{Acknowledgements}

This work was partially supported by the National Science and Technology Council, Taiwan, under Grant Numbers NSTC~107-2628-M-006-001-MY4, NSTC~111-2112-M-006-033, NSTC~111- 2119-M-007-007, and NSTC~112-2112-M-006-029.

\appendix

\section{Incapable process description}\label{ipd}

\subsection{$\chi_{\mathcal{I},\rm cre}$}\label{chicre}

An incapable process of creating entanglement, denoted as $\chi_{\mathcal{I},{\rm cre}}$, should satisfy that: for all input states that are already separable, the corresponding output states must also remain separable. With this main idea, the set of constraints, denoted as, $\mathcal{D}_{\mathcal{I}}=D(\tilde{\chi}_{\mathcal{I}, \rm cre})$, for characterizing $\chi_{\mathcal{I},{\rm cre}}$ is described by~\cite{Kuo19}:
\begin{eqnarray}
\begin{gathered} 
\tilde{\chi}_{\mathcal{I},\rm cre} \ge 0,\\
\tilde{\rho}_{\rm out} = \tilde{\chi}_{\mathcal{I},\rm cre}(\rho_{\rm in}) \ge 0\quad \forall \rho_{\rm in},\\
\tilde{\rho}^{\rm PT}_{\rm out} = \tilde{\chi}_{\mathcal{I},\rm cre}(\rho_{\rm in})^{\rm PT} \ge 0\quad \forall \rho_{\rm in} \in s_{\rm sep}, 
\end{gathered}\label{chiincre}
\end{eqnarray}
where $\tilde{\chi}_{\mathcal{I}, \rm cre}$ is an unnormalized process matrix of $\chi_{\mathcal{I},{\rm cre}}$, and $s_{\rm sep}$ denotes the set of separable states.

The first and second constraints ensure that a process matrix and its output states are positive semi-definite. The third constraint is based on the positive partial transpose (PPT) criterion~\cite{Peres96,Horodecki96} and guarantees that if the input states are separable states, the output states are separable states as well. That is, the input and output state density matrices remain positive after partial transpose operation, $\rho^{\rm PT}_{\rm in} \ge 0$ and $\tilde{\rho}^{\rm PT}_{\rm out} \ge 0$. 

In order to make the $\chi_{\mathcal{I},{\rm cre}}$ under the constraint set $D(\tilde{\chi}_{\mathcal{I}, \rm cre})$ explicitly representable and used for measuring quantum process capability (QPC), we utilize the quantum process tomography (QPT)~\cite{Chuang97,NielsenChuang} procedure to construct the constraint set $D(\tilde{\chi}_{\mathcal{I}, \rm cre})$ that a process matrix of incapable process should satisfy. All matrices satisfy $D(\tilde{\chi}_{\mathcal{I}, \rm cre})$ are considered as incapable process matrices in determining the quantities of $\alpha$ and $\beta$ by using semidefinite programming (SDP). We briefly describe how we construct the process matrix as follows. First, in QPT under the incapable process constraints~(\ref{chiincre}), the output states for the input separable states used in QPT: $\rho_{{\rm in},m}=\ketbra{m}{m}, \ket{m} \in \{\ket{0}, \ket{1}, \ket{+}, \ket{R}\}^{\otimes 2}$, where $\ket{+}=(1/\sqrt{2})(\ket{0}+\ket{1})$ and $\ket{R}=(1/\sqrt{2})(\ket{0}+i\ket{1})$, should be separable. Moreover, to ensure that when the input states are an identity matrix, the corresponding output states $\rho_{\rm out}$ are the same in the different decompositions of complementary bases. This requirement is used in tomographically characterizing separable output states out of incapable processes, where the outputs of Pauli matrices associated with the identity matrix in the inputs should be objectively described, independent of measurement bases. See Eq.~(26) in~Ref.~\cite{Kuo19} for detailed discussions.

\subsection{$\chi_{\mathcal{I},\rm pre}$}\label{chipre}

If a process matrix represents a process without the capability of preserving entanglement, denoted as $\chi_{\mathcal{I},\rm pre}$, the output state must be separable. We utilize the PPT criterion to represent it as: $\tilde{\rho}_{\rm out}^{\rm PT} \ge 0$. Based on this main condition, the set of constraints for incapable of entanglement preservation, denoted as $\mathcal{D}_{\mathcal{I}}=D(\tilde{\chi}_{\mathcal{I}, \rm pre})$, is specified as follows:
\begin{eqnarray}
\begin{gathered}
\tilde{\chi}_{\mathcal{I},\rm pre} \ge 0,\\
\tilde{\rho}_{\rm out} = \tilde{\chi}_{\mathcal{I},\rm pre}(\rho_{\rm in}) \ge 0,\\
\tilde{\rho}^{\rm PT}_{\rm out} = \tilde{\chi}_{\mathcal{I},\rm pre}(\rho_{\rm in})^{\rm PT} \ge 0\quad \forall \rho_{\rm in}.
\end{gathered}\label{chiinpre}
\end{eqnarray}

Furthermore, we use QPT to explicitly construct an SDP optimization parameter $\chi_{\mathcal{I},{\rm pre}}$ under the constraint set $D(\tilde{\chi}_{\mathcal{I}, \rm pre})$ for determining $\alpha$ and $\beta$. First, to satisfy the main constraint of 
$D(\tilde{\chi}_{\mathcal{I}, \rm pre})$~(\ref{chiinpre}), we choose the following 16 input states for QST~\cite{Chen21}: $\rho_{{\rm in},m}=\ketbra{m}{m}, \ket{m} \in \{\ket{00}, \ket{01}, \ket{0+}, \ket{0R}, \ket{10}, \ket{11}, \ket{1+}, \ket{1R}, \ket{+0}, \ket{+1},\\ \ket{R0}, \ket{R1},\ket{\phi^+}, \ket{\phi^{+i}}, \ket{\psi^+}, \ket{\psi^{+i}}\}$, where $|-\rangle=(1/\sqrt{2})(|0\rangle-|1\rangle)$, $\ket{L}=(1/\sqrt{2})(\ket{0}-i\ket{1})$, $\ket{\phi^+}=(1/\sqrt{2})(\ket{00}+\ket{11})$, $\ket{\phi^{+i}}=(1/\sqrt{2})(\ket{00}+i\ket{11})$, $\ket{\psi^+}=(1/\sqrt{2})(\ket{01}+\ket{10})$, and $\ket{\psi^{+i}}=(1/\sqrt{2})(\ket{01}+i\ket{10})$. Then, we require that the output states for these 16 input states are separable and satisfy $\tilde{\rho}^{\rm PT}_{\rm out}\geq0$. These output states after normalization $\rho_{\rm out}$ are sufficient to describe a process $\chi_{\mathcal{I},\rm pre}$ that makes all input states separable as required to satisfy the last constraint in Eq.~(\ref{chiinpre}). See Sec.~IIC. and Eq.~(A10) in Ref.~\cite{Chen21} for the detailed discussion.

\section{Measuring the quantum process capabilities}\label{mqpc}

\subsection{Composition: $\alpha_{\rm cre}$ and $\alpha_{\rm pre}$}\label{alpha}

With the constructed process matrix of incapable process as shown in Sec.~\ref{chicre}, any experimental process can be quantitatively decomposed into a linear combination of capable and incapable processes by:
\begin{eqnarray}\label{3.3}
\begin{aligned}
\chi_{\rm expt} = a\chi_{\mathcal{C},\rm cre} + (1 - a)\chi_{\mathcal{I},\rm cre},
\end{aligned}
\end{eqnarray}
where $a\ge 0$ represents the composition ratio of the two processes. The entanglement creation capability can be defined as
\begin{eqnarray}\label{3.4}
\begin{aligned}
\alpha_{\rm cre} \equiv \mathop{\min}_{\chi_{\mathcal{I},\rm cre}}a,
\end{aligned}
\end{eqnarray}
representing the minimum component of capable process that can be found in the experimental process. It can be determined through SDP with MATLAB~\cite{Lofberg04,Toh}:
\begin{eqnarray}
\begin{aligned}
\alpha_{\rm cre} = \mathop{\min}_{\tilde{\chi}_{\mathcal{I},\rm cre}}[1 - {\rm tr}(\tilde{\chi}_{\mathcal{I},\rm cre})],
\end{aligned}
\end{eqnarray}
where ${\rm tr}(\tilde{\chi}_{\mathcal{I},{\rm cre}}) = {\rm tr}((1-a)\chi_{\mathcal{I},{\rm cre}}) = 1 - a$, ${\rm tr}(\tilde{\chi}_{\mathcal{C},{\rm cre}}) = {\rm tr}(a\chi_{\mathcal{C},{\rm cre}}) = a$, and $\chi_{\rm expt} - \tilde{\chi}_{\mathcal{I},\rm cre} = \tilde{\chi}_{\mathcal{C},\rm cre} \ge 0$.

When using the incapable process matrix constructed in Sec.~\ref{chipre}, the composition of entanglement preservation capability in $\chi_{\rm expt}$: $\alpha_{\rm pre}$, can be determined in the same manner as shown above.

\subsection{Robustness: $\beta_{\rm cre}$ and $\beta_{\rm pre}$}\label{beta}

An experimental process $\chi_{\rm expt}$ can become incapable by adding noise such that
\begin{eqnarray}
\begin{aligned}
\dfrac{\chi_{\rm expt} + b\chi'}{1 + b} = \chi_{\mathcal{I},\rm cre},
\end{aligned}
\end{eqnarray}
where $b \ge 0$ and $\chi'$ represents a noise process of positive operator. The robustness of entanglement creation capability is defined as the minimum amount of noise added to the experimental process such that $\chi_{\rm expt}$ becomes $\chi_{\mathcal{I},\rm cre}$:
\begin{eqnarray}
\begin{aligned}
\beta_{\rm cre} \equiv \mathop{\min}_{\chi'}b,
\end{aligned}
\end{eqnarray}
which can solve by using SDP by
\begin{eqnarray}
\begin{aligned}
\beta_{\rm cre} = \mathop{\min}_{\tilde{\chi}_{\mathcal{I},\rm cre}}[{\rm tr}(\tilde{\chi}_{\mathcal{I},\rm cre}) - 1],
\end{aligned}
\end{eqnarray}
where  ${\rm tr}(\tilde{\chi}_{\mathcal{I},{\rm cre}}) = {\rm tr}((1+b)\chi_{\mathcal{I},{\rm cre}}) = 1 + b$ and $\chi_{\mathcal{I},{\rm cre}}$ is constructed according to the method discussed in Sec.~\ref{chicre}.

The robustness of entanglement preservation capability of $\chi_{\rm expt}$: $\beta_{\rm pre}$, can be determined in the same manner as shown above, where the incapable process matrix is constructed as described in Sec.~\ref{chipre},

\subsection{Fidelity criterion}

The fidelity criterion measures the similarity between the experimental process $\chi_{\rm expt}$ and the target process of photon fusion $\chi_{\text{fusion}}$, which can be evaluated using process fidelity $F_{\rm expt} \equiv {\rm tr}(\chi_{\rm expt}\chi_{\text{fusion}})$. It also indicates that if the experimental process exceeds the maximum imitation value of the incapable process, it is considered capable and cannot be mimicry by any incapable process, i.e.,
\begin{eqnarray}
\begin{aligned}
F_{\rm expt} \ge F_\mathcal{I} \equiv	\mathop{\max}_{\chi_{\mathcal{I},{\rm cre}}}[{\rm tr}(\chi_{\rm \mathcal{I},cre (pre)}\chi_{\text{fusion}})],
\end{aligned}
\end{eqnarray}
where $\chi_{\rm \mathcal{I},cre (pre)}$ is specified in Sec.~\ref{chicre} (\ref{chipre}) under the condition $D(\tilde{\chi}_{\mathcal{I}, \rm cre (pre)})$, Eq.~(\ref{chiincre}) [Eq.~(\ref{chiinpre})], with ${\rm tr}(\tilde{\chi}_{\mathcal{I},\rm cre (pre)}) = 1$.

\subsection{Relationhsip between $\alpha_{\rm cre}$ and $\alpha_{\rm pre}$}

$\alpha_{\rm cre}$ and $\alpha_{\rm pre}$ can be used to quantify the part of the process that has no entanglement creation but has entanglement preservation. Firstly, according to the definition of capability composition, $\chi_{\rm expt}$ can be decomposed into
\begin{eqnarray}\label{r1}
\begin{aligned}
\chi_{\rm expt} = \alpha_{\rm cre}\chi_{\mathcal{C},\rm cre} + (1 - \alpha_{\rm cre})\chi_{\mathcal{I},\rm cre},
\end{aligned}
\end{eqnarray}
where $(1 - \alpha_{\rm cre})\chi_{\mathcal{I},\rm cre}$ represents the portion that cannot generate entanglement from the separable state in the process. To quantify the part that $\chi_{\rm expt}$ can preserve entanglement but cannot generate entanglement, we decompose it using the composition of entanglement preservation capability. This decomposition represents as
\begin{eqnarray}
(1 - \alpha_{\rm cre})\chi_{\mathcal{I},\rm cre} = \alpha_{\rm pre'}\chi_{\mathcal{C},\rm pre} + (1 - \alpha_{\rm cre} - \alpha_{\rm pre'})\chi_{\mathcal{I},\rm pre}.\nonumber \\ \label{r2}
\end{eqnarray}
Therefore, by Eqs.~(\ref{r1}) and~(\ref{r2}), the process $\chi_{\rm expt}$ consists of three mutually exclusive parts:
\begin{eqnarray}\label{r3}
\chi_{\rm expt} = \alpha_{\rm cre}\chi_{\mathcal{C},\rm cre} + \alpha_{\rm pre'}\chi_{\mathcal{C},\rm pre} + (1 - \alpha_{\rm cre} - \alpha_{\rm pre'})\chi_{\mathcal{I},\rm pre},\nonumber\\
\end{eqnarray}
where $\alpha_{\rm cre}$ represents the composition that can generate entanglement, $\alpha_{\rm pre'}$ represents the part that cannot generate entanglement but can preserve entanglement, and $1 - \alpha_{\rm cre} - \alpha_{\rm pre'}$ is the part that can neither generate nor preserve entanglement. From Sec.~\ref{alpha}, $\chi_{\rm expt}$ can be expressed as
\begin{eqnarray}
\begin{aligned}
\chi_{\rm expt} = \alpha_{\rm pre}\chi_{\mathcal{C},\rm pre} + (1 - \alpha_{\rm pre})\chi_{\mathcal{I},\rm pre}.
\end{aligned}
\end{eqnarray}
From the above two equations, we can observe the following relationship between the compositions: $\alpha_{\rm pre} = \alpha_{\rm cre} + \alpha_{\rm pre'}$.
That is, the portion that cannot generate entanglement but can preserve entanglement is quantified by:
\begin{equation}
\alpha_{\rm pre'} = \alpha_{\rm pre}-\alpha_{\rm cre}.
\end{equation}

\section{Estimating the quantum process capabilities}\label{eqpc}


We will demonstrate, given a set of experimental probabilities, denoted by $\{\mathcal{P}\}$, how to find an estimated process matrix, denoted by $\chi_\text{SDP}$, and its $\alpha$ and $\beta$ via SDP under $\mathcal{D}_{\mathcal{I}}$ and the constraints of the given probabilities $\{\mathcal{P}\}$. Here $\{\mathcal{P}\}=\{F^{\rm expt}_{i \rightarrow j}\}$ is the set consisting the measured basis probabilities in experiments $F^{\rm expt}_{i \rightarrow j}$ used for estimating the lower bound of process fidelity, $F_{{\rm LB}}$~\cite{Hofmann05,Hofmann06,Okamoto08}, in our demonstration. To estimate the process capability by evaluating the information obtained from experiments, we demand that the $\chi_{\rm SDP}$ satisfies the following constraints in SDP:
\begin{eqnarray}
\begin{gathered}\label{est}
\chi_{\rm SDP} \ge 0,\\
{\rm tr}(\chi_{\rm SDP}) = 1,\\
\sum_{l,k}\sum_{m,n}\bra{j_l}(\chi_{\rm SDP})_{m,n}{E^{(2)}_m}\ketbra{i_k}{i_k}E^{(2)\dagger}_n\ket{j_l} = F^{\rm expt}_{i \rightarrow j},
\end{gathered}
\end{eqnarray}
where $E^{(2)}_{m} = E_a \otimes E_b$ for $a, b \in$ $\{ 1,2,3,4 \}$, $m = 4(a-1)+b$, and $E_1 = |0\rangle\langle 0|,$ $E_2 = |0\rangle\langle 1|,$ $E_3 = |1\rangle\langle 0|,$ $E_4 = |1\rangle\langle 1|$. Moreover, $i_k$ and $j_l$ represent 
the $k$th and $l$th results in the $i$ and $l$ measurement bases, respectively.

In the constraint set~(\ref{est}), the first requirement for the parameterized process matrix $\chi_{\rm SDP}$ must be positive semi-definite. The second constraint is that $\chi_{\rm SDP}$ should be normalized. The last constraint is that when the input is in a given specific state, $\ketbra{i_k}{i_k}$, the output state are measured on the specified basis states $\ket{j_l}$, and the classical fidelity obtained via the $\chi_{\rm SDP}$ should be equal to the value measured in the experiment, $F^{\rm exp}_{i \rightarrow j}$. This condition ensures that the given constraints are satisfied within a finite number of measurements. Explicitly, these three classical fidelities under the SDP constraints are shown as follows:
\begin{widetext}
\begin{align}\label{lf}
F^{\rm expt}_{Z \rightarrow Z} 
\!&=\!\dfrac{1}{2}(\sum_{m,n}\!{\bra{HH}\!(\chi_{\rm SDP})_{m,n}E_m^{(2)}(\ketbra{HH}{HH})E_n^{(2)\dagger}}\!\ket{HH}+\!\! \sum_{m,n}\!{\bra{VV}\!(\chi_{\rm SDP})_{m,n}E^{(2)}_m(\ketbra{VV}{VV})E_n^{(2)\dagger}}\!\ket{VV}),  \\
F^{\rm expt}_{X \rightarrow X} 
\!&=\!\dfrac{1}{4}(\sum_{m,n}\!{\bra{++}\!(\chi_{\rm SDP})_{m,n}E^{(2)}_m(\ketbra{++}{++})E_n^{(2)\dagger}}\!\ket{++} \!\!+\!\! \sum_{m,n}\!{\bra{--}\!(\chi_{\rm SDP})_{m,n}E^{(2)}_m(\ketbra{++}{++})E_n^{(2)\dagger}}\!\ket{--} \nonumber \\ \nonumber
+&\sum_{m,n}\!{\bra{+-}\!(\chi_{\rm SDP})_{m,n}E^{(2)}_m(\ketbra{+-}{+-})E_n^{(2)\dagger}}\!\ket{+-} \!\!+\!\! \sum_{m,n}\!{\bra{-+}\!(\chi_{\rm SDP})_{m,n}E^{(2)}_m(\ketbra{+-}{+-})E_n^{(2)\dagger}}\!\ket{-+}\\ \nonumber
+&\sum_{m,n}\!{\bra{+-}\!(\chi_{\rm SDP})_{m,n}E^{(2)}_m(\ketbra{-+}{-+})E_n^{(2)\dagger}}\!\ket{+-} \!\!+\!\! \sum_{m,n}\!{\bra{-+}\!(\chi_{\rm SDP})_{m,n}E^{(2)}_m(\ketbra{-+}{-+})E_n^{(2)\dagger}}\!\ket{-+}\\
+&\sum_{m,n}\!{\bra{++}\!(\chi_{\rm SDP})_{m,n}E^{(2)}_m(\ketbra{--}{--})E_n^{(2)\dagger}}\!\ket{++} \!\!+\!\! \sum_{m,n}\!{\bra{--}\!(\chi_{\rm SDP})_{m,n}E^{(2)}_m(\ketbra{--}{--})E_n^{(2)\dagger}}\!\ket{--}),\\ \nonumber
F^{\rm expt}_{X \rightarrow Y} 
\!&=\!\dfrac{1}{4}(\sum_{m,n}\!{\bra{RL}\!(\chi_{\rm SDP})_{m,n}E^{(2)}_m(\ketbra{++}{++})E_n^{(2)\dagger}}\!\ket{RL} \!\!+\!\! \sum_{m,n}\!{\bra{LR}\!(\chi_{\rm SDP})_{m,n}E^{(2)}_m(\ketbra{++}{++})E_n^{(2)\dagger}}\!\ket{LR}\\ \nonumber
+&\sum_{m,n}\!{\bra{RR}\!(\chi_{\rm SDP})_{m,n}E^{(2)}_m(\ketbra{+-}{+-})E_n^{(2)\dagger}}\!\ket{RR} \!\!+\!\! \sum_{m,n}\!{\bra{LL}\!(\chi_{\rm SDP})_{m,n}E^{(2)}_m(\ketbra{+-}{+-})E_n^{(2)\dagger}}\!\ket{LL}\\ \nonumber
+&\sum_{m,n}\!{\bra{RR}\!(\chi_{\rm SDP})_{m,n}E^{(2)}_m(\ketbra{-+}{-+})E_n^{(2)\dagger}}\!\ket{RR} \!\!+\!\! \sum_{m,n}\!{\bra{LL}\!(\chi_{\rm SDP})_{m,n}E^{(2)}_m(\ketbra{-+}{-+})E_n^{(2)\dagger}}\!\ket{LL}\\
+&\sum_{m,n}\!{\bra{RL}\!(\chi_{\rm SDP})_{m,n}E^{(2)}_m(\ketbra{--}{--})E_n^{(2)\dagger}}\!\ket{RL} \!\!+\!\! \sum_{m,n}\!{\bra{LR}\!(\chi_{\rm SDP})_{m,n}E^{(2)}_m(\ketbra{--}{--})E_n^{(2)\dagger}}\!\ket{LR}).
\end{align}
\end{widetext}

In addition to the process fidelity lower bound:
\begin{eqnarray}
F_{{\rm LB}}&=&\frac{1}{2}(F^{\rm expt}_{Z \rightarrow Z} +F^{\rm expt}_{X \rightarrow X} +F^{\rm expt}_{X \rightarrow Y} -1)\nonumber\\
&=&{\rm tr}(\chi_{\text{SDP}}\chi_{\text{fusion}}),
\end{eqnarray}
under the constraint set~(\ref{est}), we are able to evaluate the QCP of the $\chi_{\rm SDP}$ in terms of $\alpha$ and $\beta$ as shown in Sec.~\ref{alpha} and Sec.~\ref{beta}:
\begin{equation}
\chi_{\rm SDP} = a\chi_{\mathcal{C}} - (1-a)\chi_{\mathcal{I}},
\end{equation}
$\alpha \equiv \mathop{\min}_{\chi_{\mathcal{I}}}a$, and
\begin{equation}
\dfrac{\chi_{\rm SDP} + b\chi'}{1+b} = \chi_{\mathcal{I}},
\end{equation}
$\beta \equiv \mathop{\min}_{\chi'}b$.


\begin{figure*}[t]
\includegraphics[width=15cm]{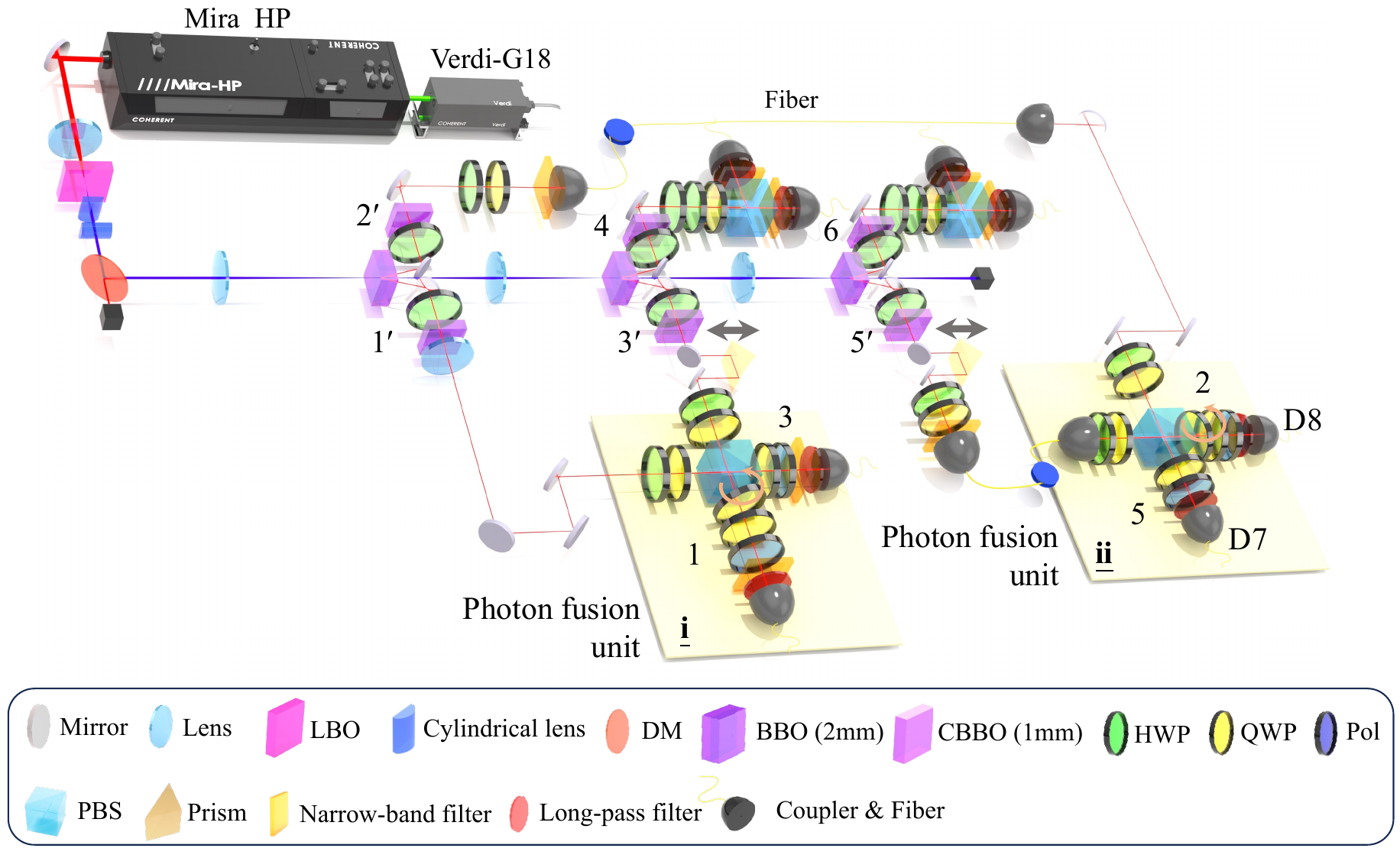}
\caption{Full experimental setup of entangling six photons with two units of photon fusion.}
\label{setup}
\end{figure*}

As detailed in the main text, it totally requires $576$ measurements of the probabilities of input-output-state events for tomographical analysis. Compared to this, the given base probability set $\{\mathcal{P}\}=\{F^{\rm expt}_{i \rightarrow j}\}$ consists of only $40$ probabilities. To prove this reduced number, first, it is clear that $F^{\rm exp}_{Z \rightarrow Z}$ considers two input states, where the output states are required to be measured in the basis corresponding to 4 events: $\{HH,HV,VH,VV\}$. There are $2\cdot 4=8$ base probabilities in $F^{\rm exp}_{Z \rightarrow Z}$. Similarly, for $F^{\rm exp}_{X \rightarrow X} $ and $F^{\rm exp}_{X \rightarrow Y}$, each includes $4\cdot4=16$ probabilities. Therefore, we have total $4+2\cdot16=40$ basis probabilities in the set $\{F^{\rm expt}_{i \rightarrow j}\}$ used to construct a $\chi_{\rm SDP}$ and to estimate the QPCs.

\section{Entangled photon sources}~\label{eps}

\subsection{The 1st and 2nd entangled photon pairs}

Figure~\ref{setup} shows the full schematic of our laser system and multi-photon experimental setup. First, we used a diode-pumped solid-state CW laser Verdi-G18 to pump a high-power ultrafast Ti:sapphire oscillator Mira HP. An ultrafast infrared pulsed laser beam was generated with an output power of about 2.6 W when Verdi-G18 operated at 13.5 W. The ultrafast infrared pulsed laser was focused on the up-converted crystal $\rm{LiB_{3}O_{5}}$ (LBO) to generate ultraviolet pulses via up-conversion. The ultraviolet laser beam was shaped to be circularized by the horizontal and vertical cylindrical lens. Then, seven dichroic mirrors (DM) filtered out the infrared light and extracted the ultraviolet light. The resulting ultraviolet laser beam has a center wavelength of 390 nm, a pulse duration of 144 fs, a repetition rate of 76 MHz, and an average power of $\sim$1.0 W.
 
 The ultraviolet pulse laser beam was focused onto three BBO crystals using focusing lenses with a $10$~cm focal length to generate entangled photons via spontaneous parametric down-conversion (SPDC) process~\cite{Kwiat95} with high brightness. including the compensation for the walk-off effect by an additional correction BBO (CBBO) crystal. The powers of the pump beam of the first and second BBOs were $\sim1.15$~W and $\sim1.0$~W, respectively. After collecting with fiber couplers, we used the silicon avalanche photodiode single-photon detector modules to measure the number of photons each detector receives, and via coincidence logic processing unit (field programmable gate array based coincidence unit) to measure the coincidence counts. In front of each fiber coupler, an narrow-band filter (central wavelength $\sim780$~nm, FWHM$\sim$3  nm, transmission rate $\ge99\%$) increases the spectral indistinguishable of the photons, facilitating our observation of the Hong-Ou-Mandel-type effect. Moreover, scattered light from the ultraviolet pump light is eliminated using a long-pass filter (460 nm$\sim$1200 nm transmission and transmission rate $\ge80\%$). Half wave plate (HWP) and quarter wave plate (QWP) are used to prepare different inputs for process tomography in each photon fusion unit. A QWP and a polarizer (Pol.) is combined for projective measurements. The relative phase between the two-photon amplitudes in a photon fusion unit is adjueted by properly rotating an additional QWP around its vertically-oriented fast axis 

\begin{table*}
\renewcommand\arraystretch{1.5}
\caption{\label{tab:table3} Summary of the created three SPDC entangled photon pairs and their related qualities. In investigating genuine four-photon EPR steering discussed in Sec.~\ref{Quantification}, the target four-photon GHZ type states are: $\left|G_4\right\rangle=(1/\sqrt{2})(\left|H_1V_{2'}H_3V_4\right\rangle+\left|V_1H_{2'}V_3H_4\right\rangle)$, with respect to the experimental creation via the first photon fusion unit on Pair 1 and Pair 2, and $\left|G_4\right\rangle=(1/\sqrt{2})(\left|H_2V_{1'}H_5V_6\right\rangle+\left|V_2H_{1'}V_5H_6\right\rangle)$, with respect to the experimental creation via the second photon fusion unit on Pair 1 and Pair 3.}
\begin{ruledtabular}
\begin{tabular}{cccc}
 &{\rm Pair 2}&{\rm Pair 1}&{\rm Pair 3}\\ \hline
 {\rm Two-fold coincidence rate (Hz)}&$95 \times 10^3$&$108 \times 10^3$ &$55 \times 10^3$ \\
{\rm Pump power (W)}&$\sim 1.0$&$\sim 1.15$&$\sim 0.85$\\
  {\rm State fidelity}&$(94.54\pm 0.08)\%$&$(92.23\pm 0.09)\%$ &$(92.21\pm 0.13)\%$\\
\multirow{2}{*}{{\rm Four-fold coincidence rate (Hz)}}&\multicolumn{2}{c}{$\sim 64$}&$-$\\
 &$-$&\multicolumn{2}{c}{$\sim 31$}\\
 \multirow{2}{*}{{\rm $|{G_4}\rangle$ state fidelity}}&\multicolumn{2}{c}{$(75.30\pm 0.79)\%$\footnote{It was measured by using the minimum-five-observable method~\cite{Kao23}.}/$(75.79\pm 0.37)\%$\footnote{It was obtained by measuring the Pauli observables~\cite{Lu20}.}}&$-$\\
 &$-$&\multicolumn{2}{c}{$(71.91\pm 0.49)\%$}\\
  \multirow{2}{*}{{\rm Fusion process fidelity}}&\multicolumn{2}{c}{$(81.55\pm 0.76)\%$}&$-$\\
 &$-$&\multicolumn{2}{c}{$(75.231\pm 0.24)\%$}\\
\end{tabular}
\end{ruledtabular}\label{3pairdata}
\end{table*}

These configurations and settings created the first two pairs of SPDC entangled photons according to the target state of $\left|\psi^-\right\rangle=(1/\sqrt{2})(\left|HV\right\rangle-\left|VH\right\rangle)$. The photon pair created in modes $1'$-$2'$ is called the first pair, and the created pairs of photons in modes $3'$-4, and $5'$-6, are called the second and third pairs, respectively. To assess the prepared state's proximity to the target state, we employ a set of measuring wave plates consisting of a  QWP and a HWP to determine its state fidelity. The state fidelity is defined as $F_{s} = {\rm tr}(\rho_{\rm expt}\ket{\psi^-}\!\!\bra{\psi^-})$, where $\rho_{\rm expt}$ is the experimental two-photon state. We determined the state fidelity by measuring the following three expectation values of observables: $\langle{X}{X}\rangle$, $\langle{Y}{Y}\rangle$ and $\langle{Z}{Z}\rangle$, where $\langle{M_1}{M_2}\rangle$ = ${\rm tr}(\rho_{\rm expt}{M_1}\otimes{M_2})$, and $M_{1,2}\in\{X,Y,Z\}$ are the Pauli operators. Experimentally, we define these operators as: $X=\ketbra{+}{+}-\ketbra{-}{-}$, $Y=\ketbra{R}{R}-\ketbra{L}{L}$, and $Z=\ketbra{H}{H}-\ketbra{V}{V}$. We observed photon pairs of $\sim108\times10^3$ per second from the first BBO, while photon pairs of $\sim95\times10^3$ per second from the second BBO. The fidelity of the first entangled photon pair is $\sim(92.23\pm0.09)\%$, and the fidelity of the second entangled photon pair is $\sim(94.54\pm0.08)\%$.

\subsection{The 3rd entangled photon pair}

The third pair of SPDC entangled photons in modes $5'$-6 was generated by directing the pump power of $390$~nm pump beam to pass through the first two BBOs to refocus on the third BBO as depicted in Fig.~\ref{setup}. The third BBO's pump power was approximately $0.85$~W. The resulting state fidelity is of $\sim(92.21\pm0.13)\%$ with respect to the target state $\ket{\psi^-}$ with a creation rate of $\sim55\times10^3$ pairs per second.

\subsection{Fiber-based photon transmission for \\photon fusion}

For the space limit of our optical table, we used fiber-based photon transmission for fusing the photons in the photon fusion unit \textbf{ii}. After completing the steps of creating all three pairs, we guided the original mode $2'$ and mode $5'$ to the unit \textbf{ii} via single-mode optical fibers, where the two optical fibers are at the same length of 2 m~\cite{Kim21}. We aligned photons in modes $2'$ and $5'$ before the PBS as parallel light to ensure consistent beam divergence and convergence behaviors. Moreover, as shown in Fig.~\ref{setup}, for the unequal lengths of free space paths of the beams of the photons generated from the first and third BBOs, three mirrors are introduced in mode 2$'$ before the PBS to make the two beam sizes and path lengths as similar as possible.

Then, we optimize the coupling efficiency of D7 and D8 to the entangled photons from mode $2'$ and mode $5'$, respectively, ensuring spatial overlapping in the interferometer. However, due to the two-fold coupling of photons through the two optical fibers, the yield of the two pairs of entangled photons decreases, and the coupling efficiency decreases accordingly.

\subsection{The experimental imperfection of photon fusion}\label{homimf}
According to theoretical predictions, when two photons overlap both temporally and spatially on PBS, one should expect to observe a four-fold coincident count rate to be the corresponding decrease. The theoretical visibility (The visibility is defined by $(N_{\rm max}-N_{\rm min})/(N_{\rm max}+N_{\rm min})$, where $N_{\rm max}(N_{\rm min})$ denotes the number of four-fold desired (undesired) coincidence
counts.) is expected to reach 100$\%$. Unfortunately, the measured observations fall short of this expectation, although they still exhibit non-classical results. Several factors contribute to this discrepancy:

(a) High-order photon pair emission. The platform utilized in our experiment is based on photon pairs generated by a femtosecond pulse laser through SPDC. In order to enhance the collection efficiency of the six-fold coincidence counting, it is necessary to intensify the pump source. However, this operation increases the effects of higher-order photon pair emission. The theoretical n-pair generation probability can be represented as $p_n$=${(n+1){\rm tanh}^{2n}r}/{{\rm cosh}^4 r}$~\cite{Kok2000}, where r is a real-valued coupling coefficient.  For our experimental results, at a two-fold coincident count rate of approximately $108\times 10^3/{\rm sec}$. The probability of creating one(two) photon pairs is $\sim$0.0987(0.0082) per pulse, it can seen that $p_1$ is 12 times larger than $p_2$. If the pump power increases and the two-fold coincident counts rate reaches approximately $130\times 10^3/{\rm sec}$. The creating probability $p_1(p_2)$ is 0.1188(0.0122) per pulse, resulting in an approximately 10 times difference~\cite{Wagenknecht10}. For two-fold coincidence counts, if the detector receives photons from different pairs, these signals will lead to an increase in undesired counts. For four-fold coincidence counts, multiple pair emissions can cause two photons from one incident direction to generate signals in the interferometer simultaneously, rather than from post-selection of fusion. These signals then contribute to the background noise, resulting in decreased visibility.

(b) Spectrum mismatch between signal and idler photons. The photons generated through the method used in our experiment exhibit spectral differences between the signal and idler photons~\cite{Yao2012}. Our solution to mitigate the visibility decrease caused by this effect is to add a narrow-band filter in front of the detectors. Testing with a narrow-band filter with a specification of FWHM = 3.5 nm yielded a measured visibility $\sim$0.62 with a $|G_4\rangle$ state fidelity $\sim70.01\%$ for the photon fusion unit $\textbf{i}$~\cite{TJT2022}. However, employing a narrow-band filter with a FWHM = 3.5 nm increased visibility $\sim$ 0.71 for the photon fusion unit $\textbf{i}$ and with a fidelity $\sim75.30\%$. The fidelity measurements for the four-photon states mentioned above were conducted using the minimum five-observable method~\cite{Kao23}.

(c) The dispersion effect of photons in optical fiber transmission leads to pulse broadening. When dispersion occurs, the group velocity and phase velocity are not equal, and their difference varies with wavelength. Since a single pulse is composed of multiple wavelength components, they travel at different group velocities through a dispersive medium. This difference in arrival times of the light signals leads to pulse broadening. Due to the use of single-mode optical fibers to guide entangled photon pairs in the second interferometer setup of this study, dispersion effects occur. Optical fibers, being dispersive media, cause pulse broadening in photons. Experimental evidence suggests this pulse-broadening phenomenon can be eliminated when two photons experience identical dispersion conditions. However, precise control over dispersion conditions comes at a cost. Our experiment employed two optical fibers of equal length to eliminate most of the pulse-broadening~\cite{Kim21}. Nevertheless, a small portion (Bends in optical fibers and the difference between optical components.) of it remains, affecting the visibility of photon fusion and causing differences in visibility between the photon fusion unit $\textbf{i}$ and $\textbf{ii}$.

\subsection{Summary of the photon pairs}

Table~\ref{3pairdata} summarizes the measured two-photon and four-photon coincidence count rates (photon pair creation rates), state fidelities, and the process fidelities under the respective pump powers. The four-photon state were created and close to a GHZ type target state: $\ket{G_4}$, via a photon fusion, either \textbf{i} or \textbf{ii}. 

\section{Measurements of process matrix}\label{mpm}

\subsection{Quantum process tomography and\\ process matrix}

The concept of QPT~\cite{Chuang97,NielsenChuang} involves determining a set of unknown operation elements {$E_{i}$} for a process $\mathcal{E}$ acting on the input state $\rho_{\rm in}$. The final state can be characterized by the equation:
\begin{eqnarray}
\begin{aligned} 
\label{eq:EIEIdagger}
\mathcal{E}(\rho_{\rm in})=\sum_{i}E_{i}\rho_{\rm in} E_{i}^{\dagger},
\end{aligned}
\end{eqnarray}
where $\sum_{i}E_{i}^{\dagger}E_{i} \leq I$.
To determine the operation elements $E_i$ from experimentally measurable data, we can utilize a fixed set of known operators $\tilde{E}_i$ to represent $\mathcal{E},$ thus 
\begin{eqnarray}
\begin{aligned} 
E_{i}=\sum_{m} e_{im} \tilde{E}_{m},
\end{aligned}
\end{eqnarray}
where $e_{im} \in \mathbb{C}.$ Then $\mathcal{E}(\rho_{\rm in})$ can be represented as
\begin{eqnarray}
\begin{aligned} 
\mathcal{E}(\rho_{\rm in})=\sum_{mn}\chi_{mn}\tilde{E}_{m}\rho_{\rm in}\tilde{E}_{n}^{\dagger}.
\end{aligned}
\end{eqnarray}
Here, $\{\tilde{E}_m\}$ represents a basis for operators on the space of density matrices, and $\chi$ is a positive Hermitian matrix with elements that can be determined from the experimental data, $\chi_{mn} = \sum_{i}e_{im}e^{\ast}_{in}$. We can fully characterize $\mathcal{E}$ by $\chi$, called the process matrix.

One can follow a systematic and experimentally feasible procedure to determined the process matrix. Let us consider a two-qbit process $\mathcal{E}$ as an example. We use the following 16 input states for QPT: $\rho_{mn} = \ketbra{mn}{mn}$, for $m,n \in \{H,V,+,R\}$. The corresponding process matrix can be represented in the following form:
\begin{align}
\chi = 
	\begin{bmatrix}
	\rho_{0000} &\rho_{0001}&\rho_{0010}&\rho_{0011}\\
	\rho_{0100} &\rho_{0101}&\rho_{0110}&\rho_{0111}\\
	\rho_{1000} &\rho_{1001}&\rho_{1010}&\rho_{1011}\\
	\rho_{1100} &\rho_{1101}&\rho_{1110}&\rho_{1111}\\ 
	\end{bmatrix}. \label{chi}
\end{align}
It is worth noting that, first, for $\mathcal{E}(\rho_{mn})$ = $\rho'_{mn}$, the output states $\rho'_{mm}$ can be used to obtain $\chi$ for precisely describing $\mathcal{E}$. Moreover, the process matrix $\chi$ is experimentally obtainable by performing quantum state tomography to know the density matrix of the output state $\rho'_{mn}$. The diagonal elements of the process matrix are as shown follows: 
\begin{eqnarray}
\rho_{0000} = \rho'_{00},\quad  \rho_{0101} = \rho'_{01}, \nonumber\\
\rho_{1010} = \rho'_{10},\quad  \rho_{1111} = \rho'_{11},\label{mn1}
\end{eqnarray}
and the other matrix elements are 

\begin{widetext}
\begin{eqnarray}
\begin{aligned}
\rho_{0001} =& \rho'_{0+} + i\rho'_{0R} - \dfrac{1+i}{2}(\rho'_{00} + \rho'_{01}),\\
\rho_{0010} =& \rho'_{+0} + i\rho'_{R0} - \dfrac{1+i}{2}(\rho'_{00} + \rho'_{10}),\\
\rho_{0111} =& \rho'_{+1} + i\rho'_{R1} - \dfrac{1+i}{2}(\rho'_{01} + \rho'_{11}),\\
\rho_{1011} =& \rho'_{1+} + i\rho'_{1R} - \dfrac{1+i}{2}(\rho'_{10} + \rho'_{11}),\\
\rho_{0011} =& \rho'_{++} + i\rho'_{+R}-\dfrac{1+i}{2}(\rho'_{+0}+\rho'_{+1}) + i(\rho'_{R+}+i\rho_{RR}-\dfrac{1+i}{2}(\rho'_{R0} + \rho'_{R1}))\\
&-\dfrac{1+i}{2}(\rho'_{0+}+\rho'_{1+}+i(\rho'_{0R}+\rho'_{1R}))-\dfrac{1+i}{2}(\rho'_{00}+\rho'_{01}+\rho'_{10}+\rho'_{11}),\\
\rho_{0110} =& \rho'_{++} - i\rho'_{+R}-\dfrac{1+i}{2}(\rho'_{+0}+\rho'_{+1}) + i(\rho'_{R+}-i\rho_{RR}-\dfrac{1+i}{2}(\rho'_{R0} + \rho'_{R1}))\\
&-\dfrac{1+i}{2}(\rho'_{0+}+\rho'_{1+}-i(\rho'_{0R}+\rho'_{1R}))-\dfrac{1+i}{2}(\rho'_{00}+\rho'_{01}+\rho'_{10}+\rho'_{11}),\\
\rho_{0100} =& \rho^\dagger_{0001},\quad \rho_{1000} = \rho^\dagger_{0010},\quad \rho_{1001} = \rho^\dagger_{0110},\\
\quad \rho_{1100} =& \rho^\dagger_{0011},\quad \rho_{1101} = \rho^\dagger_{0111},\quad \rho_{1110} = \rho^\dagger_{1011}.\label{mn2}
\end{aligned}
\end{eqnarray}
\end{widetext}
It is worth noting that ${\rm tr}(\rho'_{mn})\leq1$ describes the probability of observing the state $\rho'_{mn}$. See Eq.~(\ref{rout}) below for more detailed discussion about how $\rho'_{mn}$ were determined in our experiment.


\begin{table*}
\renewcommand\arraystretch{1.5}
\caption{\label{tab:table4}Summary of the raw data for $n/N_T$ in Eq.~(\ref{3.20}) measured by using (a) Local wave plate rotation method, and (b)
RSP method in the photon fusion unit i. The ideal values of $n/N_T$ are shown in (c).}
\begin{ruledtabular}
\begin{tabular}{cccccc}
 &\multicolumn{2}{c}{{\rm (a) Local wave plate rotation method}}&\multicolumn{2}{c}{{\rm (b) RSP method}}&{\rm (c) Ideal case} \\ 
 {\rm Input states} &{\rm Coincidence counts ($n$)}&$\dfrac{n}{N_T}$& {\rm Coincidence counts ($n$)}&$\dfrac{n}{N_T}$&$\dfrac{n}{N_T}$ \\ \hline
$|{00}\rangle$\footnote{In our experiment, we choose the input state $|00\rangle$ total coincidence counts  as our $N_T$.}	&{\rm 12379}	&{\rm 1}			&{\rm 12379}	&{\rm 1}&{\rm 1}\\
$|{01}\rangle$	&{\rm 1583}	&{\rm 0.127878}	&{\rm 1583}	&{\rm 0.127878}&{\rm 0}\\
$|{0+}\rangle$	&{\rm 7173}	&{\rm 0.579449}	&{\rm 5782}	&{\rm 0.467081}&{\rm 0.5}\\
$|{0R}\rangle$	&{\rm 5526}	&{\rm 0.446401}	&{\rm 6080}	&{\rm 0.491154}&{\rm 0.5}\\

$|{10}\rangle$	&{\rm 1785}	&{\rm 0.144196}	&{\rm 1785}	&{\rm 0.144196}&{\rm 0}\\
$|{11}\rangle$	&{\rm 11897}	&{\rm 0.961063}	&{\rm 11897}	&{\rm 0.961063}&{\rm 1}\\
$|{1+}\rangle$	&{\rm 6592}	&{\rm 0.532515}	&{\rm 6954}	&{\rm 0.561758}&{\rm 0.5}\\
$|{1R}\rangle$	&{\rm 6763}&{\rm 0.546328}	&{\rm 6738}	&{\rm 0.544309}&{\rm 0.5}\\

$|{+0}\rangle$	&{\rm 7073}	&{\rm 0.571371}	&{\rm 6456}	&{\rm 0.521528}&{\rm 0.5}\\
$|{+1}\rangle$	&{\rm 6236}	&{\rm 0.503756}	&{\rm 6862}	&{\rm 0.554326}&{\rm 0.5}\\
$|{++}\rangle$	&{\rm 5451}	&{\rm 0.440343}	&{\rm 6473}	&{\rm 0.522902}&{\rm 0.5}\\
$|{+R}\rangle$	&{\rm 6359}	&{\rm 0.513693}	&{\rm 6644}	&{\rm 0.536715}&{\rm 0.5}\\

$|{R0}\rangle$	&{\rm 6168}	&{\rm 0.498263}	&{\rm 6920}	&{\rm 0.559011}&{\rm 0.5}\\
$|{R1}\rangle$	&{\rm 5932}	&{\rm 0.479199}	&{\rm 6050}	&{\rm 0.488731}&{\rm 0.5}\\
$|{R+}\rangle$	&{\rm 5665}	&{\rm 0.457630}	&{\rm 6462}	&{\rm 0.522013}&{\rm 0.5}\\
$|{RR}\rangle$	&{\rm 6107}	&{\rm 0.493335}	&{\rm 6485}	&{\rm 0.523871}&{\rm 0.5}\\
 
\end{tabular}
\end{ruledtabular}\label{tab2}
\end{table*}

\subsection{Experimental determination of process matrix}

\subsubsection{Fusion input state preparation}

First, to prepare the necessary input states for QPT of the interference process, we employed two methods: the remote state preparation (RSP)~\cite{Pati00,Bennett01} and the local wave plate rotation methods.

The RSP method involves using one of the entangled photons (trigger) to be measured in a specific basis for preparing the state of the other photon (singal). The corresponding rotations must be applied on the signal sides to prepare the input state for process tomography. On the other hand, the local wave plate rotation method refers to directly utilizing the original state $\ket{\psi^-}$ of the entangled state in the $H/V$ basis. Similarly, through the RSP method, we still prepare the initial state at the input side of fusion unit (signal) on a $Z$ $(H/V)$ basis, but no rotation is required at the trigger side; only the signal input photon needs to be rotated via wave plates.

The use of RSP introduces its own entanglement quality error (imperfection) and transmission error during state preparation. Whereas, the method of local wave plates used to rotate the post-measurement state of entangled state in the $H/V$ basis, which yields a higher-quality signal input state.

\subsubsection{Fusion output state determination}\label{state_out}

After preparing the required input states for QPT, we consider how to determine the corresponding output states. Due to the interference process of photon fusion being non-trace preserving, when constructing the process matrix using the measured output state, we include the probability of detecting two-photon coincidence counts of photons in both of PBS output ports. The output states can be expressed as: 
\begin{eqnarray}\label{3.20}
\begin{aligned}
\mathcal{E}(\rho_{\rm in}) = \dfrac{n}{N_T}\rho_{\rm out}.\label{rout}
\end{aligned}
\end{eqnarray}
where $N_T$ denotes the total photon pair number at the input of PBS for a specific input state $\rho_{\rm in}$, and $n$ represents the total number of coincidences detected at the two output ports of PBS for the normalized $\rho_{\rm out}$. Here $\rho_{\rm out}$ is determined by quantum state tomography. Compared to Eqs.~(\ref{chi})-(\ref{mn2}), we have used $\rho_{\rm in}=\rho_{mn}$ and $\rho'_{mn}=(n/N_T)\rho_{\rm out}$.

In our experiments, we aim that the two input photons are initially prepared in the state of $\ket{H}$. Then we use local wave plates to transform $\ket{H}$ to required states for QPT. Therefore, given $\ket{HH}$ as inputs, we can consider the the total coincidence counts of $\ket{HH}$ detected at the two output ports of PBS as $N_T$. See Table.~\ref{tab2} for the raw data measured in the photon fusion init \textbf{i}.

\subsubsection{Maximum likelihood estimation}\label{ml}

Our tomographic data are used to obtain a completely
positive physical process matrix by finding a positive Hermitian matrix that is the closest fit in a least-squares sense. Following the approaches introduced in Refs.~\cite{James01,OBrien04}, this is achieved by writing a Hermitian parametrization $\vec{t}$ of the experimental process matrix and minimizing a likelihood function $f(\vec{t})$ so that the observed data is most probable to constitute a physical process matrix that is completely positive under the assumed statistical and physical model. Then, we obtain the maximum likelihood estimation of the experimental process matrix. Therefore, the obtained $\alpha$ and $\beta$ describe the quantum process capabilities under the maximum likelihood estimation for the experimental process matrices.

The likelihood function is in the form
\begin{eqnarray}
f(\vec{t})\!\!&=&\!\!\sum_{i,j,k,l}
\!\frac{1}{\mathcal{N}}\!\!\left(\!\!n_{i_k j_l}\!\!-\!\!
\mathcal{N}\sum_{m,n}\!\bra{j_l}\!\tilde{\chi}_{mn}(\vec{t}){E^{(2)}_m}\!\ketbra{i_k}{i_k}\!E^{(2)\dagger}_n\!\!\ket{j_l}
\!\!\right)^2 \nonumber\\
&&+\hspace{3pt}\lambda\left(\sum_{m,n,k} \tilde{\chi}_{mn}(\vec{t})\hspace{3.5pt}\!\! {\rm Tr}\left(E^{(2)\dagger}_n E^{(2)}_k E^{(2)}_m\!\right)\! -\! \delta_{k,0}\!\!\right)\!,
\end{eqnarray}
where $n_{i_k j_l}$ is the measured number of coincident counts for a given specific input $\rho_{\rm in}=\ketbra{i_k}{i_k}$ and the measurement is on the specified basis states $\ket{j_l}$, $\mathcal{N}$ is the total number of coincident photon pairs within the counting time, $\lambda$ is a weighting factor that can be adjusted to ensure that the matrix is arbitrarily close to a completely positive map and $\delta$ is the Kronecker delta. The first sum on the right fits the data to a Hermitian matrix via a least-squares fit, and the second sum enforces the set of further constraints for the orthogonality of $E^{(2)}_m$. Due to the interference process of photon fusion being non-trace preserving ($\sum_{i}E_{i}^{(2)\dagger}E_{i}^{(2)} <I$), an additional constraint ${\rm Tr}\left(\sum_{m,n}\tilde{\chi}_{mn}(\vec{t}){E^{(2)}_m}\ketbra{i_k}{i_k}E^{(2)\dagger}_n\right)=n_{i_k}/N_T\hspace{3pt}, \forall i_k$, discussed in Appendix~\ref{state_out} is required in maximum likelihood estimation, where $n_{i_k}=\sum_{j_l}n_{i_k j_l}$ represents the total number of coincidences detected at the two output ports of PBS for a given specific input $\rho_{\rm in}=\ketbra{i_k}{i_k}$ and $N_T$ denotes the total photon pair number for a specific input state $\ketbra{i_k}{i_k}$ as shown in Table.~\ref{tab:table4}. (We choose the counts of input state $|00\rangle$ as our total coincidence counts $N_T$ in our experiment.)

\begin{figure*}[t]
\includegraphics[width=14cm]{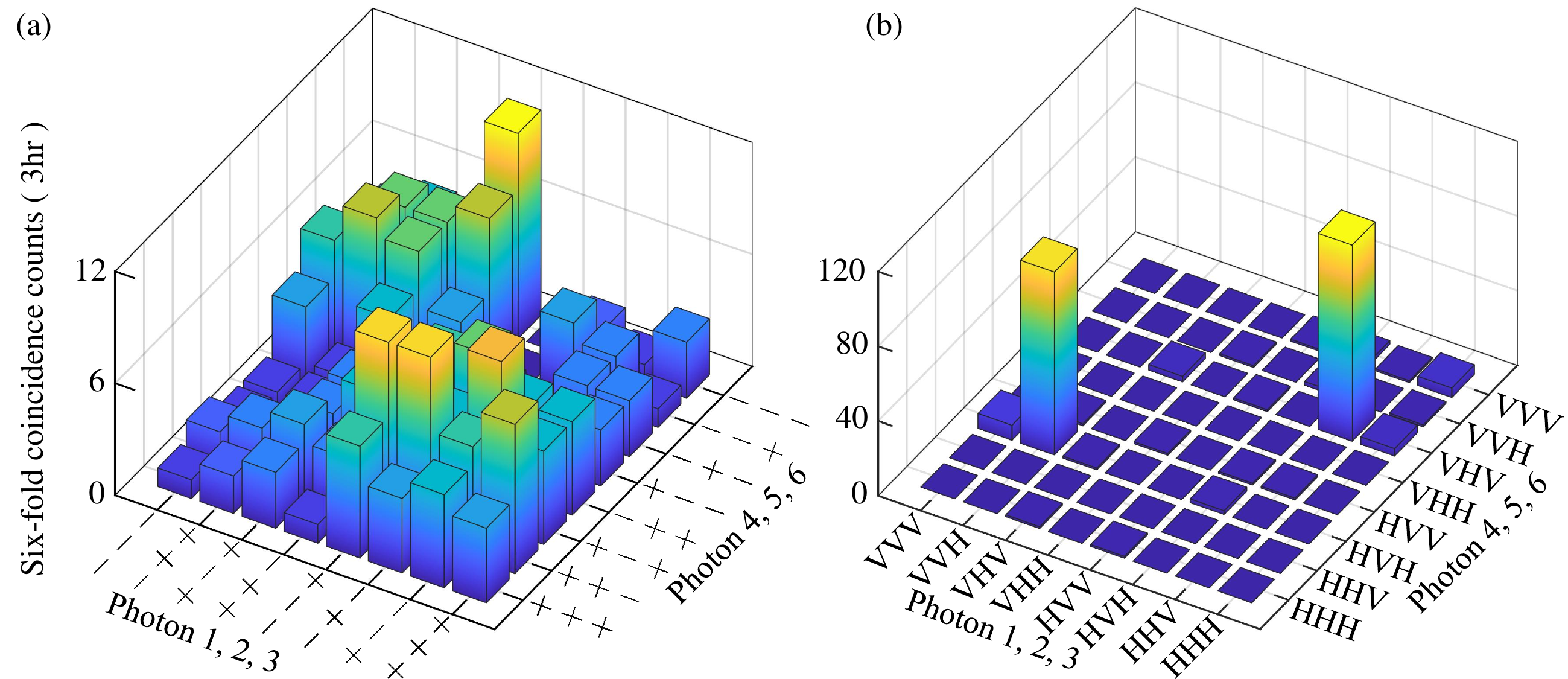}
\caption{The result of entanglement witness measurement. (a) Six-fold coincidence counts under the measurement setting of $X^{\otimes 6}$. (b) Six-fold coincidence counts under the measurement setting of $Z^{\otimes 6}$. The data collection time is 3 hours for these two measurements.}
\label{witness6photon}
\end{figure*}

\section{Six-photon state fidelity}\label{spsf}

In order to determine the six-photon state fidelity directly by using the QPC quantifications of the experimental photon fusion units, we consider the state mapping:
\begin{equation}
\tilde{\rho}_{\rm out}=\chi_{\text{expt,\textbf{ii}}}\chi_{\text{expt,\textbf{i}}}(\rho_{1'2'}\otimes\rho_{3'4}\otimes\rho_{5'6}),
\end{equation}
where $\chi_{\text{expt,\textbf{ii}}}$ and $\chi_{\text{expt,\textbf{i}}}$ have been experimentally determined. The created entangled photon pairs are described by using the Werner-type state with a type-II down-conversion dephasing noise by 
\begin{eqnarray}\label{C.2}
&&\rho_{Wi} = p_{2i-1}\dfrac{I\otimes I}{4} + \dfrac{p_{2i}}{2}(\ketbra{HV}{HV} + \ketbra{VH}{VH} ) \nonumber\\
&&\ \ \ \ \ \ \ \ + (1 -p_{2i-1} - p_{2i})\ketbra{\psi^-}{\psi^-},
\end{eqnarray}
for $i=1,2,3$, where $p_{2i-1}$ and $p_{2i}$ represent noise intensities for the white noise and the dephasing noise, respectively. Furthermore, we optimize the noise intensities $p_{2i-1}$ and $p_{2i}$ by comparing the two-photon states $\rho_{W1}$, $\rho_{W2}$ and $\rho_{W3}$, with the experimentally obtained density matrices, $\rho_{1'2'}$, $\rho_{3'4}$, and $\rho_{5'6}$, respectively, to have the best state fidelity~\cite{NielsenChuang} by
\begin{equation}
F_{s,i} = \sqrt{\sqrt{\rho_{\rm expt}}\rho_{Wi}\sqrt{\rho_{\rm expt}}}.\nonumber
\end{equation}
After optimization, we obtain
\begin{eqnarray}
F_{s,1}&=&0.9716,\nonumber\\
F_{s,2}&=&0.9744,\\
F_{s,3}&=&0.9677.\nonumber
\end{eqnarray}
Therefore, as $\rho_{\rm in} = \rho_{W1}\otimes\rho_{W2}\otimes\rho_{W3}$, the corresponding output state $\tilde{\rho}_{\rm out}$ can be explicitly obtained.
There are 108 different terms of six-photon state components in the output state. With the values of $\alpha_{\textbf{i}}$, $\alpha_{\textbf{ii}}$, and $p_i$ for $i=1,2,...,5$, we then arrive at a fidelity of
\begin{equation}
F_s=\bra{G_6}\rho_{G_6}\ket{G_6}=0.5755,\label{fs}
\end{equation}
for the estimated six-photon state $\rho_{G_6}=\tilde{\rho}_{\rm out}/{\rm tr}(\tilde{\rho}_{\rm out})$, where 
\begin{equation}
\left|G_6\right\rangle=\frac{1}{\sqrt{2}}(\left|H_1V_2H_3V_4V_5H_6\right\rangle+\left|V_1H_2V_3H_4H_5V_6\right\rangle),
\end{equation}
is the target state in our experiment. Since $F_s$ is more significant than $1/2$ that any biseparable states can achieve~\cite{Bourennane04}, the created state $\rho_{G_6}$ is genuinely six-photon entangled. 

\section{Witness to genuine six-pohton entanglement}\label{wspe}

In the previous section, we estimate the state fidelity of the created six-photon entanglement. Here, to support this result, we determine its characteristic of genuine multipartite entanglement by using the information obtained from the actual measurement in the experiment. 

In general, entanglement witness~\cite{Guhne09} is the most common method to detect multipartite entanglement. If the expectation value of a witness operator, $\mathcal{W}_{\rm GHZ_6}$, is less than zero, ${\rm tr}(\mathcal{W}_{\rm GHZ_6}\rho_{{\rm GHZ}_6}) < 0$, then the experimental state $\rho_{{\rm GHZ}_6}$ possesses genuine multipartite entanglement. This witness operator is designed for detecting states close to the target six-qubit GHZ state of $\ket{\text{GHZ}_6}=(1/\sqrt{2})(\ket{H}^{\otimes6}+\ket{V}^{\otimes6})$ and is represented as follows~\cite{Toth05}:
\begin{eqnarray}\label{4.5c}
\begin{aligned}
\mathcal{W}_{\mathrm{GHZ}_{6}}&:=3I-2[\dfrac{S_{1}^{\mathrm{GHZ}_{6}}+I}{2}+\prod_{k=2}^{6}\dfrac{S_{k}^{\mathrm{GHZ}_{6}}+I}{2}],\\
\end{aligned}\label{witnessghz1}
\end{eqnarray}
where $I$ is the identity operator,
\begin{eqnarray}
\begin{aligned}
S_{1}^{\mathrm{GHZ}_{6}}&=\bigotimes_{k=1}^{6}X_{k},\\
S_{k}^{\mathrm{GHZ}_{6}}&=Z_{k-1}Z_{k},\ \mathrm{for}\,k=2, 3, \ldots, 6,
\end{aligned}\label{witnessghz2}
\end{eqnarray}
$X_k=X$, and $Z_k=Z$. Here, only two observable measurements, $X^{\otimes 6}$ and $Z^{\otimes 6}$, are sufficient to determine the value of the entanglement witness.

Since the target state $\left|G_6\right\rangle$ in our experiment is equivalent to the state $\ket{\text{GHZ}_6}$, we utilized a witness kernel, denoted $\mathcal{W}_{\rm G_6}$, which is obtained by replacing $Z_k$ by $-Z_k$ for $k=2,4,5$ in Eq.~(\ref{witnessghz2}) for the witness kernel~(\ref{witnessghz1}) to detect genuine six-photon entanglement. Figure~\ref{witness6photon} describes the measured six-fold coincidence counts obtained from the experiments under two measurement settings of $X^{\otimes 6}$ and $Z^{\otimes 6}$. The average counting rate is approximately 250 photons every 3~hours (0.025/s). With the measured six-fold coincidence counts, we determined all the expectation values of the observables used in the witness kernel. For example, the expectation values for the $\langle X^{\otimes 6}\rangle$ and $\langle Z^{\otimes 6}\rangle$ observables are $0.4745\pm 0.054$ and $0.6047\pm 0.041$, respectively. Based on these expectation values, we obtain the entanglement witness of our six-photon GHZ state is
\begin{equation}
{\rm tr}(\mathcal{W}_{\rm G_6}\rho_{G_6}) = -0.0801\pm 0.037.
\end{equation}
This indicates that the state $\rho_{G_6}$ we have generated possess genuine six-photon entanglement. This is also consistent with the result examined by using the criterion based on the estimated state fidelity~(\ref{fs}).


\begin{thebibliography}{12}

\bibitem{Pan98} J.-W. Pan and A. Zeilinger, Greenberger-Horne-Zeilinger-state analyzer, Phys. Rev. A \textbf{57}, 2208 (1998).

\bibitem{Pan01} J.-W. Pan, C. Simon, $\check{{\rm C}}$.  Brukner, and A. Zeilinger, Entanglement purification for quantum communication, Nature \textbf{410}, 1067 (2001). 

\bibitem{Pan03} J.-W. Pan, S. Gasparoni, R. Ursin, G. Weihs, and A. Zeilinger, Experimental entanglement purification of arbitrary unknown states, Nature \textbf{423}, 417 (2003).

\bibitem{Browne05} D. E. Browne and T. Rudolph, Resource-efficient linear optical quantum computation, Phys. Rev. Lett. \textbf{95}, 010501 (2005).

\bibitem{Bodiya06} T. P. Bodiya and L.-M. Duan, Scalable generation of graph-state entanglement through realistic linear optics. Phys. Rev. Lett. \textbf{97}, 143601 (2006).

\bibitem{Lu07} C.-Y. Lu, X.-Q. Zhou, O. G{\"u}hne, W.-B. Gao, J. Zhang, Z.-S. Yuan, A. Goebel, T. Yang, and J.-W. Pan, Experimental entanglement of six photons in graph states, Nature Phys \textbf{3}, 91 (2007).

\bibitem{Pan12} J.-W. Pan, Z.-B. Chen, C.-Y. Lu, H. Weinfurter, A. Zeilinger, and M. {\.Z}ukowski, Multiphoton entanglement and interferometry, Rev. Mod. Phys. \textbf{84}, 777 (2012).

\bibitem{Wang16} X.-L. Wang, L.-K. Chen, W. Li, H.-L. Huang, C. Liu, C. Chen, Y.-H. Luo, Z.-E. Su, D. Wu, Z.-D. Li, H. Lu, Y. Hu,
X. Jiang, C.-Z. Peng, L. Li, N.-L. Liu, Y.-A. Chen, C.-Y. Lu,
and J.-W. Pan, Experimental Ten-Photon Entanglement, Phys. Rev. Lett. \textbf{117}, 210502 (2016).

\bibitem{Zhong18} H.-S. Zhong, Y. Li, W. Li, L.-C. Peng, Z.-E. Su, Y. Hu, Y.-M. He, X. Ding, W. Zhang, H. Li, L. Zhang, Z. Wang, L.-X. You, X.-L. Wang, X. Jiang, L. Li, Y.-A. Chen, N.-L. Liu, C.-Y. Lu, and J.-W. Pan, 12-Photon Entanglement and Scalable Scattershot Boson Sampling with Optimal Entangled-Photon Pairs from Parametric Down-Conversion, Phys. Rev. Lett. \textbf{121}, 250505 (2018).

\bibitem{Adcock19} J. C. Adcock, C. Vigliar, R. Santagati, J. W. Silverstone, and M. G. Thompson, Programmable four-photon graph states on a silicon chip, Nat. Commun. \textbf{10}, 3528 (2019).

\bibitem{Llewellyn20} D. Llewellyn, Y. Ding, I. I. Faruque, S. Paesani, D. Bacco, R. Santagati, Y.-J. Qian, Y. Li, Y.-F. Xiao, M. Huber, M. Malik, G. F. Sinclair, X. Zhou, K. Rottwitt, J. L. O’Brien, J. G. Rarity, Q. Gong, L. K. Oxenlowe, J. Wang, and M. G. Thompson, Chip-to-chip quantum teleportation and multi-photon entanglement in silicon, Nat. Phys. \textbf{16}, 148 (2020).

\bibitem{Wang20} J. Wang, F. Sciarrino, A. Laing, and M. G. Thompson, Integrated photonic quantum technologies, Nat. Photon. \textbf{14}, 273 (2020).

\bibitem{MeyerScott22} E. Meyer-Scott, N. Prasannan, I. Dhand, C. Eigner, V. Quiring, S. Barkhofen, B. Brecht, M. B. Plenio, and C. Silberhorn, Scalable Generation of Multiphoton Entangled States by Active Feed-Forward and Multiplexing, Phys. Rev. Lett. \textbf{129}, 150501 (2022).

\bibitem{Pan00} J.-W. Pan, D. Bouwmeester, M. Daniell, H. Weinfurter, and A. Zeilinger, Experimental test of quantum nonlocality in three-photon Greenberger–Horne–Zeilinger entanglement, Nature (London) \textbf{403}, 515 (2000).

\bibitem{Lu20} H. Lu, C.-Y. Huang, Z.-D. Li, X.-F. Yin, R. Zhang, T.-L. Liao, Y.-A. Chen, C.-M. Li, and J.-W. Pan, Counting Classical Nodes in Quantum Networks, Phys. Rev. Lett. \textbf{124}, 180503 (2020).

\bibitem{Wu22} D. Wu, Q. Zhao, C. Wang, L. Huang, Y.-F. Jiang, B. Bai, Y. Zhou, X.-M. Gu, F.-M. Liu, Y.-Q. Mao, Q.-C. Sun, M.-C. Chen, J. Zhang, C.-Z. Peng, X.-B. Zhu, Q. Zhang, C.-Y. Lu, and J.-W. Pan, Closing the locality and detection loopholes in multiparticle entanglement selftesting, Phys. Rev. Lett. \textbf{128}, 250401 (2022).

\bibitem{Walther05} P. Walther, K. J.  Resch, T. Rudolph, E. Schenck, H. Weinfurter, V. Vedral, M. Aspelmeyer, and A. Zeilinger, Experimental one-way quantum computing, Nature (London) \textbf{434}, 169 (2005).

\bibitem{Lu09} C.-Y. Lu, W.-B. Gao, O. G{\"u}hne, X.-Q. Zhou, Z.-B. Chen, and J.-W. Pan, Demonstrating Anyonic Fractional Statistics with a Six-Qubit Quantum Simulator, Phys. Rev. Lett. \textbf{102}, 030502 (2009).

\bibitem{Yao12} X.-C. Yao, T.-X. Wang, W.-B. G. Hao-Ze Chen, A. G. Fowler, R. Raussendorf, Z.-B. Chen, N.-L. Liu, C.-Y. Lu, Y.-J. Deng,
Y.-A. Chen, and J.-W. Pan, Experimental demonstration of topological error correction, Nature \textbf{482}, 489 (2012).

\bibitem{Kimble08} H. J. Kimble, The quantum internet, Nature \textbf{453}, 1023 (2008).

\bibitem{Ritter12} S. Ritter, C. N{\"o}lleke, C. Hahn, A. Reiserer, A. Neuzner, M. Uphoff, M. M{\"u}cke, E. Figueroa, J. Bochmann, and G. Rempe, An elementary quantum network of single atoms in optical cavities, Nature \textbf{484}, 195–200 (2012).

\bibitem{Wehner18} S. Wehner, D. Elkouss, and R. Hanson, Quantum internet: A vision for the road ahead, Science \textbf{362}, eaam9288 (2018).

\bibitem{He13} Q. Y. He and M. D. Reid, Genuine Multipartite Einstein-Podolsky-Rosen Steering, Phys. Rev. Lett. \textbf{111}, 250403 (2013).

\bibitem{Li15} C.-M. Li, K. Chen, Y.-N. Chen, Q. Zhang, Y.-A. Chen, and J.-W. Pan, Genuine High-Order Einstein-Podolsky-Rosen Steering, Phys. Rev. Lett. \textbf{115}, 010402 (2015).

\bibitem{Armstrong15} S. Armstrong, M. Wang, R. Y. Teh, Q. Gong, Q. He, J. Janousek, H.-A. Bachor, M. D. Reid, and P. K. Lam, Multipartite Einstein–Podolsky–Rosen steering and genuine tripartite entanglement with optical networks, Nat. Phys \textbf{11}, 167 (2015).

\bibitem{Cavalcanti15} D. Cavalcanti, P. Skrzypczyk, G. H.  Aguilar, R. V. Nery, P. H. Souto Ribeiro, and S. P. Walborn, Detection of entanglement in asymmetric quantum networks and multipartite quantum steering, Nat. Commun. \textbf{6}, 7941 (2015).

\bibitem{Kao23} W.-T. Kao, C.-Y. Huang, T.-J. Tsai, S.-H. Chen, S.-Y. Sun, Y.-C. Li, T.-L. Liao, C.-S. Chuu, H. Lu, and C.-M. Li, Scalable Quantum Network Determination with Einstein-Podolsky-Rosen Steering, arXiv:2303.17771.

\bibitem{Chuang97} I. L. Chuang and M. A. Nielsen, Prescription for experimental determination of the dynamics of a quantum black box, J. Mod. Opt. 44, 2455 (1997).

\bibitem{NielsenChuang} M. A. Nielsen and I. L. Chuang, \textit{Quantum Computation and Quantum Information} (Cambridge University Press, 2000).

\bibitem{Bell12} B. Bell, A. S. Clark, M. S. Tame, M. Halder, J. Fulconis, W. J. Wadsworth, and J. G. Rarity, Experimental characterization of photonic fusion using fiber sources, New J. Phys. \textbf{14} 023021 (2012).

\bibitem{Hsieh17} J.-H. Hsieh, S.-H. Chen, and C.-M. Li, Quantifying quantum- mechanical processes, Sci. Rep. \textbf{7}, 13588 (2017).

\bibitem{Kuo19} C.-C. Kuo, S.-H. Chen, W.-T. Lee, H.-M. Chen, H. Lu, and C.-M. Li, Quantum process capability, Sci. Rep. \textbf{9}, 20316 (2019).

\bibitem{Chen21} S.-H. Chen, M.-L. Ng, and C.-M. Li, Quantifying entanglement preservability of experimental processes, Phys. Rev. A \textbf{104}, 032403 (2021).

\bibitem{Hofmann05} H. F. Hofmann, Complementary Classical Fidelities as an Efficient Criterion for the Evaluation of Experimentally Realized Quantum Operations, Phys. Rev. Lett.  \textbf{94}, 160504 (2005).

\bibitem{Hofmann06} H. F. Hofmann, Analysis of an experimental quantum logic gate by complementary classical operations, Mod. Phys. Lett. A  \textbf{21}, 1837 (2006).

\bibitem{Okamoto08} R. Okamoto, J. L. O’Brien, H. F. Hofmann, T. Nagata, K. Sasaki,
and S. Takeuchi, An Entanglement Filter, Science \textbf{323}, 483 (2008).

\bibitem{Peres96} A. Peres, Separability Criterion for Density Matrices, Phys. Rev. Lett. \textbf{77}, 1413 (1996).

\bibitem{Horodecki96} M. Horodecki, P. Horodecki, and R. Horodecki, Separability of mixed states: Necessary and sufficient conditions, Phys. Lett. A \textbf{223}, 1 (1996).

\bibitem{Kwiat95} P. G. Kwiat, K. Mattle, H. Weinfurter, A. Zeilinger, A. V. Sergienko, and Y. Shih, New high-intensity source of polarization-entangled photon pairs, Phys. Rev. Lett. \textbf{75}, 4337 (1995).

\bibitem{Hong87} C. K. Hong, Z. Y. Ou, and L. Mandel, Measurement of subpicosecond time intervals between two photons by interference, Phys. Rev. Lett. \textbf{59}, 2044 (1987).

\bibitem{GHZ} D. M. Greenberger, M. A. Horne, A. Shimony, and A. Zeilinger, Bell’s theorem without inequalities, Am. J. Phys. \textbf{58}, 1131 (1990).

\bibitem{James01}D. F. V. James, P. G. Kwiat, W. J. Munro, and A. G. White, Measurement of qubits, Phys. Rev. A \textbf{64}, 052312 (2001).

\bibitem{OBrien04} J. L. O'Brien, G. J. Pryde, A. Gilchrist, D. F. V. James, N. K. Langford, T. C. Ralph, and A. G. White, Quantum Process Tomography of a Controlled-NOT Gate, Phys. Rev. Lett. \textbf{93}, 080502 (2004).

\bibitem{Bourennane04} M. Bourennane, M. Eibl, C. Kurtsiefer, S. Gaertner, H. Weinfurter, O. G{\"u}hne, P. Hyllus, D. Bru{\ss}, M. Lewenstein, and A. Sanpera, Experimental Detection of Multipartite Entanglement using Witness Operators, Phys. Rev. Lett. \textbf{92}, 087902 (2004).

\bibitem{Toth05} G. T\'oth and O. G\"uhne, Detecting Genuine Multipartite Entanglement with Two Local Measurements, Phys. Rev. Lett. \textbf{94}, 060501 (2005).

\bibitem{Huang23} W.-H. Huang, S.-H. Chen, C.-H. Chang, T.-L. Hsu, K.-J. Wang, and C.-M. Li, Quantum Correlation Generation Capability of Experimental Processes, Adv Quantum Technol. \textbf{6}, 2300113 (2023).

\bibitem{Lofberg04} J. L\"ofberg, Yalmip: A toolbox for modeling and optimization
in MATLAB. In CACSD, 2004 IEEE International Symposium
on Taipei, Taiwan). Available at http://users.isy.liu.se/johanl/
yalmip/.

\bibitem{Toh} K. C. Toh, M. J. Todd, and R. H. T\"ut\"unc\"u, SDPT3 – a MATLAB software package for semidefinite-quadratic-linear programming, version 4.0. Available at http://www.math.nus.edu.sg/mattohkc/sdpt3.html.

\bibitem{Kim21}  D.-G Im, Y. Kim and Y.-H. Kim, Dispersion cancellation in
a quantum interferometer with independent single photons, Opt. Express \textbf{29}, 2348 (2021).

\bibitem{Kok2000} P. Kok and S. Braunstein, Postselected versus nonpostselected quantum
teleportation using parametric down-conversion, Phys. Rev. A \textbf{61}, 042304 (2000).

\bibitem{Wagenknecht10} C. Wagenknecht, C.-M. Li, A. Reingruber, X.-H. Bao, A. Goebel,
Y.-A. Chen, Q. Zhang, K. Chen, and J.-W. Pan, Experimental demonstration of a heralded
entanglement source, Nat. Photonics \textbf{4}, 549-552 (2010).

\bibitem{Yao2012} X.-C. Yao, T.-X. Wang, P. Xu, H. Lu, G.-S. Pan, X.-H. Bao, C.-Z. Peng, C.-Y. Lu, Y.-A. Chen, and J.-W. Pan, Observation of eight-photon entanglement, Nat. Photonics \textbf{6(4)}, 225-228 (2012).

\bibitem{TJT2022} T.-J. Tsai, Experimental realization of semi-device-independent one-way quantum computation, Master’s thesis, National Cheng Kung University, 2022.

\bibitem{Pati00} A. K. Pati, Minimum classical bit for remote preparation and measurement of a qubit, Phys. Rev. A \textbf{63}, 014302-014306 (2000).

\bibitem{Bennett01} C. H. Bennett, D. P. DiVincenzo, P. W. Shor, J. A. Smolin, B. M. Terhal, and W. K. Wootters, Remote state preparation, Phys. Rev. Lett. \textbf{87}, 077902 (2001).

\bibitem{Guhne09} O. G\"uhne and G. T\'oth, Entanglement detection, Phys. Rep. \textbf{474}, 1 (2009).

\end{thebibliography}
\end{document}